\documentclass[sn-mathphys-num]{sn-jnl}

\usepackage[table]{xcolor}
\usepackage{graphicx}%
\usepackage{multirow}%
\usepackage{amsmath,amssymb,amsfonts}%
\usepackage{amsthm}%
\usepackage{mathrsfs}%
\usepackage[title]{appendix}%
\usepackage{xcolor}%
\usepackage{textcomp}%
\usepackage{manyfoot}%
\usepackage{booktabs}%
\usepackage{algorithm}%
\usepackage{algorithmicx}%
\usepackage{algpseudocode}%
\usepackage{listings}%

\theoremstyle{thmstyleone}%
\theoremstyle{thmstyletwo}%
\newtheorem{remark}{Remark}%

\theoremstyle{thmstylethree}%

\begin{document}

\title[Analyzing homogenous and heterogeneous multi-server queues via neural networks ]{Analyzing homogenous and heterogeneous multi-server queues via neural networks}


\author*[1,2]{\fnm{Eliran} \sur{Sherzer}}\email{scherzter@gmail.com}

\affil*[1]{\orgdiv{Industrial Engineering}, \orgname{Ariel University}, \orgaddress{\street{65 Ramat HaGolan}, \city{Ariel}, \postcode{407000},  \country{Israel}}}

\affil[2]{\orgdiv{Rotman school of manamgent}, \orgname{University of Toronto}, \orgaddress{\street{105 St. George}, \city{Toronto}, \postcode{ON Canada}}}


\abstract{In this paper, we use a machine learning approach to predict the stationary distributions of the number of customers in a single-staiton multi server system. We consider two systems, the first is $c$ homogeneous servers, namely the $GI/GI/c$ queue. The second is a two-heterogeneous server system, namely the $GI/GI_i/2$ queue. We train a neural network for these queueing models, using the first four inter-arrival and service time moments. We demonstrate empirically that using the fifth moment and beyond does not increase accuracy. 

Compared to existing methods, we show that in terms of the stationary distribution and the mean value of the number of customers in a $GI/GI/c$ queue, we are state-of-the-art. Further, we are the only ones to predict the stationary distribution of the number of customers in the system in a $GI/GI_i/2$ queue. We conduct a thorough performance evaluation to assert that our model is accurate. In most cases, we demonstrate that our error is less than 5\%. Finally, we show that making inferences is very fast, where 5000 inferences can be made in parallel within a fraction of a second. }

\keywords{$GI/GI/c$ queues, Heterogeneous servers queues, neural network}



\maketitle

\section{Introduction}\label{sec:intro}

Multi-server queues are commonly found in systems like airports, factories, and hospitals, making them a subject of considerable research interest for analyzing their dynamics and quantifying their performance.

Exact results for the multi-server queueing systems are known only for Markovian systems (e.g., \cite{SEELEN1986118, LI2016866, app14010424}). When the Markovian assumptions are relaxed, researchers move to approximation methods. For example, the $GI/GI/c$ queue has several approximation methods \cite{https://doi.org/10.1111/j.1937-5956.1993.tb00094.x, Chaves2022, doi:10.1057/jors.1988.45}. However, most approximation methods are based only on the first two moments of the inter-arrival and service time distributions. According to~\cite{Gupta2010}, no approximation based on only the first two moments can be accurate
for all job size distributions for the $M/GI/c$ queue. Therefore, the results in~\cite{Gupta2010} are likely applicable to the $GI/GI/c$ system studied in this paper, as it represents a more general and complex case compared to the $M/GI/c$ queue.

A common method for analyzing the performance of complex queueing models is through computer simulation (e.g., see \cite{doi:10.1080/07408170590899625, 10.1145/2000494.2000497}). However, as noted by \cite{doi:10.1287/opre.2016.1554}, a significant limitation of simulation-based optimization methods is the substantial computational effort required to obtain optimal solutions for service operation problems in multidimensional stochastic networks. Thus, a new approach is warranted. This paper proposes a steady-state analysis of a multi-server system using machine learning (ML).

In past years, there has been a growing number of queueing models which approximate queueing measures~\cite{10015451,  sherzer23,sherzer2024computingsteadystateprobabilitiestandem, Nii20, math13050776,SHERZER2024, doi:10.1080/00207543.2021.1887536}. These models attack different queueing problems, such as transient single-server queues or tandem queueing systems. However, all these studies except two (\cite{Nii20, math13050776}) consider only single-server stations. While~\cite{Nii20, math13050776} considers a $G/G/c$, they are also based on a two-moment approximation which yields large errors. The majority of papers use neural networks (NNs) as their ML framework~\cite {10015451, SHERZER2024, sherzer23,sherzer2024computingsteadystateprobabilitiestandem, Nii20,math13050776}, with one exception where~\cite{doi:10.1080/00207543.2021.1887536}  uses Gaussian process.

This paper examines two queueing models. The first is the $GI/GI/c$ model, where both inter-arrival and service times follow general renewal processes, and the system comprises $c$ identical servers. We aim to approximate the stationary distribution of the number of customers in the system. To demonstrate the robustness of our approach, we also consider a $GI/GI_i/2$ model with two heterogeneous servers, again focusing on approximating the stationary distribution.

We adopt a supervised learning approach, selecting NNs as the ML framework for both models. Applying supervised learning with NNs to analyze a queueing system presents three main challenges. First, generating a sufficiently diverse set of training input instances, i.e., inter-arrival and service time distributions, is necessary. The resulting dataset must provide broad coverage of the distribution space—ideally representing all possible positive-valued distributions.

Second, obtaining the output labels for a given input instance is challenging. In this context, labels correspond to the system distribution values (i.e., the steady-state distribution of the number of customers in the system). Constructing the training set requires accurately generating distributions for numerous instances of intractable queueing models.

Third, representing continuous distributions, such as inter-arrival and service time distributions, discretely, as required by NN inputs. The input must be converted into a tensor format, which can lead to information loss. These challenges were previously raised in~\cite{sherzer23}, necessitating the development of a suitable distribution-to-tensor mapping.

To address the first challenge, we utilize the Phase-Type (PH) family of distributions. This family is known to be dense within the class of all non-negative distribution functions; a PH distribution with a sufficient number of phases can approximate any non-negative distribution to arbitrary precision, see~\citep[Theorem 4.2, Chapter III]{Asmussen2003}. This property enables the construction of a diverse training set by sampling from the space of PH distributions.

However, sampling from the PH space is non-trivial, as the parameters describing a PH distribution are interdependent and must satisfy multiple constraints. To overcome this, we employ a flexible sampling method capable of generating a diverse set of distributions, which was developed in~\cite{sherzer23}.

For the second challenge, labeling the input data requires discrete event simulations. While simulation is computationally expensive, and this may seem counterproductive—given that ML is intended to reduce dependence on simulations—this limitation is mitigated because labeling occurs offline. During this phase, substantial computational resources can be temporarily allocated without the need for real-time response. Once trained, the ML model enables near real-time inference, in contrast to simulations, which can take minutes to hours. Furthermore, the ML approach only requires a finite training dataset. After training, the model can make predictions for an infinite number of instances.

To tackle the third challenge—efficiently encoding inter-arrival and service time distributions for the NN—we use the first $n$ moments of the input distributions. These moments can be computed analytically when expressions for the distributions are available or estimated from observed data. The value of $n$ is determined through hyperparameter tuning during training. For the heterogeneous $GI/GI_i/2$ model, the network is provided with the first $n$ moments of the service time distributions for both servers. 

Investigating the impact of the choice of $n$ on the model’s accuracy is of particular interest. Similar studies in~\cite{10015451, sherzer23, sherzer2024computingsteadystateprobabilitiestandem, SHERZER2024} suggest that, in stationary settings, capturing the first four to five moments is typically sufficient to represent queueing dynamics effectively. In Section~\ref{sec:mom_anal}, we conduct a similar analysis and demonstrate convergence occurs at $n = 4$.

Once we train our model, we compare our approach to other $GI/GI/c$ models ~\cite{Marchal1985, KLB76, doi:10.1057/jors.1988.45,Chaves2022, https://doi.org/10.1111/j.1937-5956.1993.tb00094.x}, as will be elaborated in the literature review in Section~\ref{sec:lit_rew}. We point out that not all approximation methods aim to estimate the steady-state probabilities as we do, but only the mean number of customers. For such models, we compare the mean value instead.  We further note that we are unaware of approximation methods for heterogeneous servers in a $GI/GI_i/2$ queue. As shown in Section~\ref{sec:result}, our model is accurate and state-of-the-art compared to existing methods.

 Our results demonstrate that the trained neural network can infer the steady-state distribution for thousands of queueing systems in a fraction of a second, making it an invaluable tool for real-time decision-making in operational settings. This remarkable speed is particularly advantageous for solving non-convex optimization problems, where conventional numerical approaches often struggle with computational feasibility. By enabling rapid brute-force exploration of the state-space, our method allows for practical optimization within realistic time constraints, unlocking new possibilities for system design, resource allocation, and performance analysis in complex queueing environments.

The main contributions of this paper are: (i) An approximation method for the steady-state probabilities of a $GI/GI/c$ queue. (ii) An approximation method for the steady-state probabilities of a $GI/GI_i/2$ queue with heterogeneous servers. (iii) An empirical analysis indicates the effect of the $n^{th}$ moment of the inter-arrival and service time distributions on the stationary queue length.  (iV) An open access package that can be used for inference for both the $GI/GI/c$ and the $GI/GI_i/2$ models ( \url{https://github.com/eliransher/Multi_server_queues.git})

The outline of this paper is as follows. In Section~\ref{sec:lit_rew}, we provide a literature review. In Section~\ref{sec:prob_form}, we formalize the problem we wish to solve. In Section~\ref{sec:overview}, we describe an overview of our solution.  In Section~\ref{sec:training_proc}, we present the training process of our ML framework. In Sections~\ref{sec:exper} and~\ref{sec:result}, we detail our experiment to validate the model and present the results, respectively. In Section~\ref{sec:num_example}, we utilize our ML approach by solving an optimization problem in a multi-server queueing system. In Section~\ref{sec:discussion}, we discuss possible extensions and limitations of our approach.  
Finally, we conclude our paper in Section~\ref{sec:conclusions}. 

\section{Literature review}\label{sec:lit_rew}

Exact analytic solutions for the general case of $GI/GI/c$ queues are unavailable; however, approximation methods have been developed over the years. One of the earliest contributions was by Marchal~\cite{Marchal1985}, who proposed a method for approximating the mean sojourn time. This work is based on previous results from \cite{doi:10.1080/05695557608975111} and~\cite{KLB76}. In~\cite {doi:10.1057/jors.1988.45}, Shore approximated not only the steady-state mean number in the system but also the entire distribution. Whitt~\cite{https://doi.org/10.1111/j.1937-5956.1993.tb00094.x} also developed equations to approximate the mean number of customers for a $GI/GI/c$ queue. A more recent paper that also approximates the mean average number of customers in the system is done by Chaves and Gosavi~\cite{Chaves2022}. The authors make an approximation designed for a specific traffic intensity between 0.5 and 0.8. All these approximations are based on only the first two moments of the inter-arrival and service time distribution, which ignores vital information of the queue dynamics~\cite{Gupta2010}. As these papers are doing a similar task to ours, in Section~\ref{sec:result} we compare our model to theirs.

Some papers take an algorithmic approach for the stationary analysis of multi-server queues. Seelen~\cite{SEELEN1986118} proposed an algorithm for the numerical analysis of multi-server queueing systems with PH interarrival and service time distributions, covering both finite and infinite capacity models. However, the size of the PH representation limits the algorithm's practicality, which can be prohibitive in certain cases. Similarly, Bertsimas \cite{doi:10.1287/opre.38.1.139} presents an algorithm specifically designed for a $C_k/C_m/c$ queue, where both interarrival and service times follow Coxian distributions. While this approach is efficient and has exponential complexity with respect to $s$ and $m$, it remains polynomial in $k$. Nonetheless, the computational cost can become significant when both 
$k$ and $m$ are large, posing practical challenges in such scenarios. Furthermore, the paper does not address the general $GI/GI/c$ system.

Other papers also dealt with multi-server systems, but not with general inter-arrival and service time distributions. For example,  Kimura in~\cite{KIMURA1995157} develops approximations for the delay probability in an $M/GI/c$ queue. Chao and Zhao~\cite{CHAO1998392} consider $GI/M/c$ queues with two classes of vacation mechanisms: Station vacation and server vacation. 

There have been only a few previous studies that employ NNs for queue analysis. The most closely related work to ours is by~\cite{Nii20}, where the authors introduce a deep learning model for analyzing a $GI/GI/c$ queue. Their model takes as input only the first two moments of the inter-arrival and service time distributions and predicts the average waiting time, but it is limited to a narrow range of parameter values. Our objectives are significantly more ambitious as we aim to capture the entire stationary distribution of the system rather than just the average waiting time. Another similar work was done in~\cite{math13050776}, which used an ML approach to study the mean number of customers in a $GI/GI/c$ queue. Just like~\cite{Nii20}, they also used only the first two moments. Neither paper published an open-access code, nor compared their results to existing approximations. Therefore, we cannot compare their results to ours.

Until now, all the studies considered have focused on systems with homogeneous servers. However, several works have addressed queueing models with heterogeneous servers. Li and Stanford~\cite{LI2016866} developed a multi-class, multi-server $M/M_i/c$ queueing model with heterogeneous servers operating under the accumulating priority queueing discipline. Boxma et al.\cite{Boxma2002} analyzed a heterogeneous $M/G/2$ queue, where service times at server 1 follow an exponential distribution, while those at server 2 follow a general distribution $B(\cdot)$. Their work provides an exact analysis of the queue length and waiting time distribution, specifically when $B(\cdot)$ has a rational Laplace-Stieltjes transform. Calvo and Arteaga~\cite{app14010424} investigated heterogeneous systems with limited capacity, where arrivals follow a Poisson process, and service times are exponentially distributed with each server operating at a different rate. However, none of these studies address the scenario in which both the inter-arrival and service time distributions are non-Markovian.

This work directly builds upon the author's previous studies. In~\cite{10015451} and~\cite{sherzer23}, the stationary distribution of the number of customers in the system was approximated for $M/G/1$ and $GI/GI/1$ systems. In~\cite{SHERZER2024}, an RNN-based approach was used to analyze a transient $G(t)/GI/1$ model. Lastly,~\cite{sherzer2024computingsteadystateprobabilitiestandem} examined a tandem queueing system with single-server stations. However, none of these studies addressed multi-server queues.

\section{Problem formulation}\label{sec:prob_form}

As mentioned above, we aim to study two systems, $GI/GI/c$, and $GI/GI_i/2$. Next, we give an exact formulation and notation for both systems. We commence with introducing basic notation. Let $\textbf{A}$ be an $ n$-size vector for the first $n$ inter-arrival moments. Let $\textbf{S}_i$ be an $ n$-size vector for the first $n$ inter-arrival moments of the $i^{th}$ service distribution. For the $GI/GI/c$, we have only one service time distribution, hence $i \in \{1\}$, however, for the  $GI/GI_i/2$, $i\in \{1,2\}$. Let $c$ be the number of servers, where in the $GI/GI_i/2$ system, $c=2$. Finally, let $\textbf{P}$,  be an $l$ size vector, where the $i^{th}$ element $P_i$ denotes the probability of having $i$ customers in the system in steady-state, where $l$ reflects a cutoff point where we truncate the number of customers in the system distirubiton. The value of $l$ is discussed in Section~\ref{sec:output_generation}. We can now conclude the infernce problems of both systems is:

\noindent \textbf{$GI/GI/c$}: We wish to approxiamte  $\textbf{P}$, via $\textbf{A}$, $\textbf{S}_1$ and $c$.

\noindent \textbf{$GI/GI_/2$}: We wish to approxiamte  $\textbf{P}$, via $\textbf{A}$, $\textbf{S}_1$ and $\textbf{S}_2$.

\section{Solution overview}\label{sec:overview}

\begin{figure}
\centering
\includegraphics[scale=0.4]{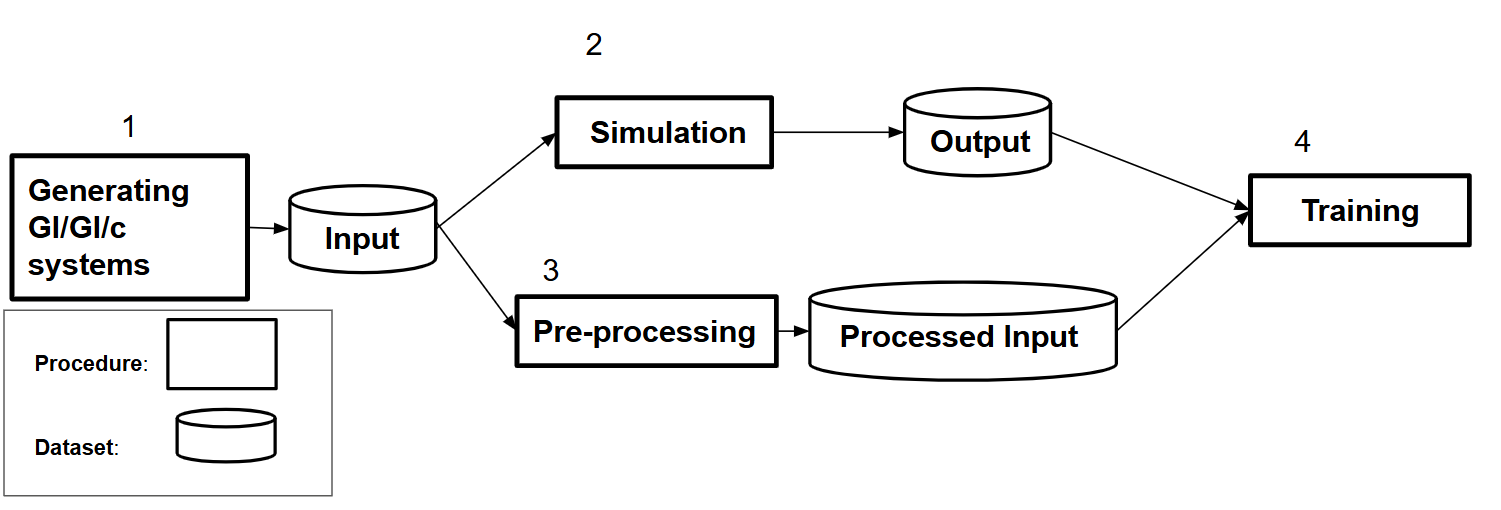}
\caption{Work-flow diagram of our learning procedure. }
\label{fig:diagram_overview}
\end{figure}

\begin{figure}
\centering
\includegraphics[scale=0.55]{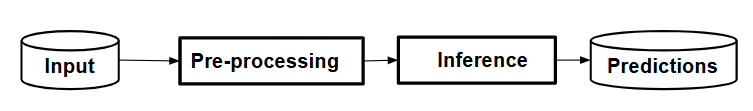}
\caption{Work-flow diagram for inference. }
\label{fig:diagram_inf}
\end{figure}

The key stages of our solution are illustrated in Figure~\ref{fig:diagram_overview}, which outlines the learning process, and Figure~\ref{fig:diagram_inf}, which depicts the inference procedure.

Steps 1–3 in Figure~
\ref{fig:diagram_overview} corresponds to the generation of training input and output data, as detailed in Section~\ref{sec:data}. The process begins with \textit{Step 1: Generating $GI/GI/c$ and $GI/GI_i/2$ systems} (described in Section~\ref{sec:input_generation}). This step produces the \textit{Input}, which consists of three components for each instance: (i) the inter-arrival time distribution, i.e., $\textbf{A}$, (ii) the service time distributions, i.e., $\textbf{S}_i$, where for the $GI/GI/c$,  $i \in \{1\}$, and for  $GI/GI_i/2$ system $i \in \{1,2 \}$.  (iii) the number of servers, i.e., $c$. 

In \textit{Step 2: Simulation}, we determine the distribution of the number of customers in the system using discrete event simulation, as explained in Section~\ref{sec:output_generation}. The \textit{Output} of this step—namely, the full steady-state distribution—is subsequently used as the target during training.

Next, in \textit{Step 3: Pre-processing}, the inter-arrival and service time distributions are processed by computing the first $n$ moments. The hyperparameter\footnote{Hyperparameters refer to model-related factors that are not learned during training, such as $n$ moments, the number of hidden layers, or the number of nodes in each layer.} $n$ requires tuning. The computed moments are then standardized, as discussed in Section~\ref{sec:pre_process}. It is important to note that this pre-processing is specific to NN training and is not used in the simulation, meaning it has no direct connection to \textit{Step 2}.

Finally, in \textit{Step 4: Training}, the processed inputs and training outputs are fed into the neural network (as described in Section~\ref{sec:network}), where the training process takes place. Once training is complete, inference follows the procedure depicted in Figure~\ref{fig:diagram_inf}. Specifically, we input the number of servers, and the inter-arrival and service time distributions, which we pre-process, and the NN produces the steady-state probabilities as its output.

\section{Training process}\label{sec:training_proc}

In this section, we first describe the data generation process (see Section~\ref{sec:data}), then we present the NN architecture (see Section~\ref{sec:network}), and finally, we present the NN loss function (see Section~\ref{sec:loss_func}).

\subsection{Data generating}\label{sec:data}

In this section, we outline the data generation process. We first present the input generation, which describes how non-negative continuous distributions are generated (see Section~\ref{sec:input_generation}).

\subsubsection{Input generation}\label{sec:input_generation}

The input consists of three parts. The first two, inter-arrival and service time distributions, are non-negative continuous distributions, and the third is the number of servers. Generating the number of servers is only applied for the $GI/GI/c$ system, where in the $GI/GI_i/2$ it must be 2. Generating $c$ is trivial, where we sample uniformly over the set $\{1,2,...,10\}$. The more challenging task is generating a wide range of distributions.

We employ the sampling technique introduced in~\cite{sherzer23}. To generate distributions, we utilize the PH family of distributions, which, as mentioned earlier, is known to be dense within the class of all non-negative distributions. Although we sample from PH distributions up to fixed order, in this case, 500, this property enables us to construct a diverse training set in practice. Indeed,~\cite{sherzer23} demonstrates extensive distribution coverage and favorably compares this method with alternative sampling techniques. The Squared Coefficient of Variation (SCV) of the sampled distributions in this study ranges from 0.0025 to 20, ensuring a versatile set of distributions.

In~\cite{sherzer23}, all sampled distributions have a mean value of 1. For inter-arrival distributions, we assume, without loss of generality, that the mean value is 1,  which hence remains unchanged. The service time distributions should be scaled to obtain a wide range of utilization levels, i.e., the aggregated fraction of time the servers are busy. We explore the range where the system utilization level $\rho$ is within $[0.001, 0.95]$.

\noindent \textbf{$GI/GI/c$ system:}  $\rho$ determines straightforwardly the service rate. This is done via the relation $\rho = \lambda/  c \mu$, where $\mu$ is the service rate and $\lambda$ is the arrival rate. To sample a $GI/GI/c$ system we follow several steps:

\begin{enumerate}
    \item Sample two distributions according to the sampling algorithm in~\cite{sherzer23}, one for inter-arrival and the other for service times.
    \item Sample $c$, from a discrete uniform distirbution over the range $[1,10]$.
    \item Sample $\rho$ from a continous uniform distirbution over the range $[0.01,0.95]$. 
    \item Derive $\mu$ the service rate such that $\rho = \lambda / c \mu$, and scale the service time accordingly\footnote{Recall the $\lambda$ is set as 1}. 
\end{enumerate}

\noindent \textbf{$GI/GI_i/2$ system:} In a heterogeneous system, given $\rho$, $\lambda$, and the number of servers (i.e., 2), the service rates cannot be determined analytically as in the $GI/GI/c$ system. Moreover, the fraction of time each server remains busy is no longer identical across all servers. In the $GI/GI/c$ system, $\rho$ represents both the fraction of time each server is busy and its average utilization. However, in the $GI/GI_i/2$ system, $\rho$ corresponds to the average fraction of time that the servers are busy.

In the $GI/GI/c$ system, once we sampled $\rho$, we were able to derive the service rate, but it is no longer the case in the $GI/GI_i/2$ system. Instead, we propose a sampling technique in which we achieve a wide range of $\rho$ by sampling the service rates.

 The central question is how to sample the service rates, denoted as $\mu_1$ and $\mu_2$, to ensure that $\rho$ lies within the range $[0.01, 0.95]$. To approach this, we treat the $GI/GI_i/2$ system as a $GI/GI/1$, where its service rate, which we denote as $\mu$, is the sum of the service rates in the $GI/GI_i/2$ system.  We first sample $\mu$. Given that the arrival rate is set to 1, we obtain $\rho = 1/\mu$. The value of $\mu$ is sampled such that the $GI/GI/1$ utilization level $\rho_{GI/GI/1}$ follows a uniform distribution over the interval $[0.01, 0.95]$, i.e., $1/\mu \sim U(0.01, 0.95)$.

Returning to the $GI/GI_i/2$ system, we define $\mu = \mu_1 + \mu_2$, where $\mu_1 \sim U(\phi, \mu - \phi)$ and $\mu_2 = \mu - \mu_1$, with $\phi$ representing the minimum permissible value for $\mu_1$ and $\mu_2$. To prevent situations where one server is almost always busy while the other effectively behaves as a $GI/GI/1$ queue, a lower bound of $\phi = 0.01$ is imposed.

Since $\mu = \mu_1 + \mu_2$ is sampled to ensure that $\rho \in [0.01, 0.95]$ in the $GI/GI/1$ framework, this does not necessarily imply that the same range applies in the $GI/GI_i/2$ system. However, empirical results suggest that $\rho \in [0.005, 0.955]$ effectively captures the intended range, which is slightly broader than the initially targeted interval. Nevertheless, it achieves our goal. 

It is crucial to note that $\rho$ is obtained through simulation, as it cannot be directly derived from $\mu_1$ and $\mu_2$, and it does not equal $\rho_{GI/GI/1}$. The procedure for generating the $GI/GI_i/2$ system is outlined below:

\begin{enumerate}
    \item Sample three distributions according to the~\cite{sherzer23}, one for inter-arrival and two  for service time distributions.
    \item Sample $1/\mu \sim U(0.01,0.95)$. 
    \item Sample, $\mu_1 \sim U(\phi, \mu - \phi)$ and compute $\mu_2$ such that $\mu_2  = \mu - \mu_1 $. 
    \item Scale the service time according to $\mu_1$ and $\mu_2$.  
\end{enumerate}

\subsubsection{Pre-processing Input }\label{sec:pre_process}

Once \textit{Step 1} is complete, these inputs undergo two pre-processing steps, as described below.

First, we analytically compute the first $n$ moments of the inter-arrival and service time distributions. For the second step, to mitigate the wide range of moment values, which may burden the learning process, we apply a log transformation to the remaining moments. This follows common practices in NN models to ensure that input values fall within a similar range. Notably, as the order of moments increases, their values grow exponentially, meaning that even standardizing the moments does not effectively constrain their range and adds computational complexity. Through experimentation, we find that the log transformation is both simpler to apply and more effective in reducing the variability across all inter-arrival and service time moments. This also enjoys an emprical succuess from previuos studies~\cite{10015451, sherzer23, SHERZER2024, sherzer2024computingsteadystateprobabilitiestandem}.

\subsubsection{Output generation}\label{sec:output_generation}

To label our data, we perform simulations. For each instance, we simulate 60 million arrivals. The simulations are written in Python via the package \textit{Simpy}. The desired output is a vector that describes the steady-state probabilities of having $i \geq 0$ customers in the system. In theory, this should be an infinite-size vector. In practice, we must provide the computed distribution to the NN as a finite and fixed-dimensional vector. Thus, we truncate our computation at $l$ such that the total probability of having more than $l$ customers is smaller than $\delta$ (both $l$ and $\delta$ are hyper-parameters). 
Under the set-up of $\rho \leq 0.95$, in our empirical evaluation, we used $\delta=10^{-3}$ to achieve the desired performance with $l=500$ in all generated samples. This corresponds to covering the probability of between $0$ and $500$ customers in the system. 

\begin{remark}
    Although the training instance is restricted to $\rho \leq 0.95$, inference can be done for any $\rho$. If there is a positive probability of having more than $l$ customers, this would decrease the accuracy of our prediction. Of course, this method is not limited to $l = 500$, where larger values can be used by simply changing the number of nodes in the output layer (see Section~\ref{sec:network}). For practical reasons, $l = 500$ was chosen as typically queues don't amount to more than 500. 
\end{remark}

To assess the accuracy of our simulations, we conduct experiments on 500 $GI/GI/c$ systems and 500 $GI/GI_i/2$ systems to evaluate labeling consistency. Each system is simulated ten times, including 60 million arrivals. We then compute the length of the 95\% confidence interval (CI) for the mean number of customers in the system. The average CI lengths for the $GI/GI/c$ and $GI/GI_i/2$ systems are 0.0132 and 0.0182, respectively. 

The average time a simulation with 60 million arrivals takes is 1.64 hours, using an Intel Xeon Gold 5115 Tray Processor with 128GB RAM. 

\subsubsection{Segmentation of the test set by SCV, $\rho$ and number of servers}\label{sec:segmentation}

To validate the model's accuracy, we construct diverse test sets. As common in queueing models (\cite{https://doi.org/10.1111/j.1937-5956.1993.tb00094.x, https://doi.org/10.1002/nav.22010}), those test sets are partitioned into groups of utilization levels $\rho$ and SCV values of the inter-arrival and service time distributions.  For this purpose, we consider four utilization level intervals ($[0.01,0.25]$, $[0.25,0.5]$,$[0.5,0.75]$,$[0.75,0.95]$), and two groups of SCV, below and above 1. In the $GI/GI/c$ system, we have two distributions (i.e., inter-arrival and service time), hence two SCV values to account for. The number of servers is partitioned into two groups, $[1,5]$ and $[6,10]$. As a result, the results are split into $32$ different groups. 

In the $GI/GI_i/2$,  we maintain the same four utilization level groups but have three SCV values to account for (i.e., one inter-arrival and two service times distributions). Of course, only one group of servers, as it is constant at 2. Hence, we remain with 32 groups.

\subsubsection{Datasets: training, validation and test}\label{sec:datasets}

To train our NN model, we require two different datasets: training (used to tune network parameters) and validation (used to tune the hyper-parameters). Both are generated using procedures described in Sections \ref{sec:input_generation} and~\ref{sec:output_generation} above. For both the  $GI/GI/c$ and the $GI/GI_i/2$ systems, the training and validation sizes are 500,000 and 50,000, respectively.  

When it comes to evaluating the performance of our model, we wanted to make sure this evaluation was not limited only to datasets generated in the same way as our training and validation data. Thus, our test sets were composed of two types of instances. 

 \noindent \textbf{Test set (i):} instances generated using the same procedure as our training and validation data. As mentioned in Section~\ref{sec:segmentation}, the test set is split into 32 groups.  Hence, we have 1,500 samples for each group, which sum to 48,000  instances. To illustrate the diversity of Test set (i), in Appendix~\ref{append:scv_rho_testset1} we plotted histograms of the inter-arrival and service distributions SCV (see Figures~\ref{fig:low_scv} and~\ref{fig:high_scv} ), and the utilization levels (see Figure~\ref{fig:rhos}). This test set is valuable due to its complexity of inter-arrival and service time distribution (e.g., a wide range of SCV). Also, as demonstrated in~\cite{sherzer23}, the generating method outperforms existing PH generating procedures concerning diversity. 
 \\ \noindent \textbf{Test set (ii):} An external data set based on an experiment conducted by~\cite{You19}, where the inter-arrival and service time follow Log-Normal, Gamma, Erlang, Hyper-Exponential, and Exponential distributions. The idea is to demonstrate that our ML predictions perform well for instances outside our sampling algorithm (which also considers non-PH distributions). 

We next summarize the experiment settings in~\cite{You19}, which created Test set (ii). Their experiments were not originally designed for multi-level queues; instead, they were for single-station queueing networks. Yet, this setting is considered an acceptable benchmark and has been used in several queueing papers (see \cite{https://doi.org/10.1002/nav.22010, doi:10.1287/opre.2015.1367, sherzer23}). 

The test set is constructed out of the following distributions:

\begin{enumerate}
    \item Exponential ($M$) distribution with mean $1/\lambda$ and SCV = 1.
    \item Erlang ($E_k$) distribution with mean $1/ \lambda$, SCV$= 1/k$, i.e., the summation of $k$
    i.i.d. exponential random variables, each with mean $1/(\lambda k)$.
    \item Hyperexponential ($H_2(SCV)$) distribution, i.e., a mixture of two exponential distributions that can be parameterized by its first three moments or the mean $1/\lambda$, SCV, and
    the ratio between the two components of the mean $r = p_1/\lambda_1 /(p_1/\lambda_1 + p_2/\lambda_2)$ where
    $\lambda_1 > \lambda_2$. Only the case $r = 0.5$ is considered.
    \item Log-normal ($LN(SCV)$) distribution.
    \item Gamma ($G(4)$) distribution with SCV = $4$.
\end{enumerate}

For the inter-arrival time distribution, the following cases are considered: $E_4$, $LN(0.25)$,
$H_2(4)$, $LN(4)$  and $G(4)$. For service-time distribution, the exponential ($M$) distribution is also considered. The arrival rate was fixed
at 1. For each combination of the 5 inter-arrival distributions and 6 service-time distributions,
20 utilization levels were considered $\rho \in \{0.01,0.06,..,0.96\}$. In the 
$GI/GI/c$ system, we also consider 10 different values of a number of servers $\{1,2,..10 \}$. We summarize the sizes of Test set (ii) under the $GI/GI/c$ and $GI/GI_i/2$ systems in Table~\ref{tab:test2_nums}. This is aligned with the segmentation of the SCV, $\rho$, and the number of servers as presented in Section~\ref{sec:segmentation}. 

\begin{table}[!htp]\centering
\caption{Test set (ii) sizes for  $GI/GI/c$ and $GI/GI_i/2$.} \label{tab:test2_nums}
\scriptsize
\begin{tabular}{|l|r|r|r|r|r|r|}\toprule
&\# Inter-arrival &\# Service &\# Servers &\# $\rho$ &Total size \\ \hline
$GI/GI/c$ &5 &6 &10 &20 &6000 \\ \hline
&\# Inter-arrival &\# Service 1 &\# Service 2 & &Total \\ \hline
$GI/GI_i/2$ &5 &6 &6 &20 &3600 \\ \hline
\bottomrule
\end{tabular}
\end{table}

\begin{remark}
    Our model requires that the first $n$ moments of the inter-arrival and service time distributions be finite, as infinite values are not suitable inputs. While some theoretical cases involving heavy-tailed distributions may lack finite moments and thus fall outside the scope of our model, such cases are uncommon in practice. Empirical evidence supports this: in~\cite{sherzer23}, the authors examined extensive real-world data from large call centers and two major hospitals over several years. Their analysis covered more than 2,500 interarrival and service time distributions each, all of which exhibited only finite moments.
\end{remark}

\subsection{Network Architecture}\label{sec:network}

We employ a Fully-Connected feed-forward NN consisting of six hidden layers, with an increasing number of neurons in each layer: (50, 70, 200, 350, 200, 350, 600, 500), resulting in a total of 599820 parameters. We used the exact same architecture and sizes for both the $GI/GI/c$ and $GI/GI_i/2$ systems. While alternative architectures such as Convolutional Neural Networks (CNNs) and Recurrent Neural Networks (RNNs) were considered, neither was deemed suitable for this task. Specifically, CNNs are not well suited due to the small size of the input vector, while RNNs are more appropriate for time-dependent systems rather than stationary ones; see \cite{NIPS2015_b618c321} for further discussion.  

For all hidden layers, we use the Rectified Linear Unit (ReLU) as the activation function, defined as 

\[
ReLU(\text{input}) = \max\{\text{input}, 0\}.
\]

Additionally, we apply the Softmax function to the output layer, defined as  

\[
SOFTMAX_i ([a_1, \dots, a_n]) = \frac{\exp(a_i)}{\sum_{j=1}^n \exp(a_j)}, \quad i=1, \ldots, n,
\]

which ensures that the output-layer weights sum to 1.

\subsection{Loss function}\label{sec:loss_func}

Our training data is divided into batches of size $B$ (a hyper-parameter). As mentioned, the stationary system distribution (output label) corresponding to each training instance is truncated to size $l$.

For a given batch, let $Y$ and $\hat{Y}$ denote matrices of size $(B, l)$, representing the true stationary system distribution values and the NN predictions, respectively. Our loss function $Loss(\cdot)$ is given by
\begin{equation}  \label{eq:loss}
     Loss(Y,\hat{Y}) =   \frac{1}{B}\sum_{i=1}^{B}\sum_{j=0}^{l-1}|Y_{i,j}-\hat{Y}_{i,j}|+ 
     \frac{1}{B}\sum_{i=1}^{B}max_j(|Y_{i,j}-
     \hat{Y}_{i,j}|),
\end{equation}

\noindent where $Y_{i,j}$ and $\hat{Y}_{i,j}$ correspond to the $i^{th}$ sample in the batch and the probability of having $j$ customers in the system for the true and predicted distributions, respectively. The batch size $B$ is tuned during training, while the output size $l$ was set to $500$, as mentioned earlier. We follow the same loss function as originally suggested by~\cite{sherzer23}. For further explanation of why this loss function was formed, see Section 5.3 in~\cite{sherzer23}.

\section{Experiments}\label{sec:exper}

In this section, we first present accuracy metrics we use to validate our model as done in Section~\ref{sec:acc_matircs}. This is followed by Section~\ref{sec:mom_anal}, where we describe how we examine the effect of the $i^{th}$ moment on the accuracy, determining how many moments to use in our model. Finally, in Section~\ref{sec:peva}, we detail how our model accuracy is being tested and how it is compared against other models. 

\subsection{Accuracy metrics}\label{sec:acc_matircs}

Our evaluation metrics are used for two purposes: one for fine-tuning the model and the other for accuracy evaluation. Next, we present the metrics, which only the first is used for fine-tuning. All the suggested methods are widely used in queueing experiments (see~\cite{sherzer23, sherzer2024approximatinggtgi1queuesdeep, You19, Marchal1985, doi:10.1057/jors.1988.45, Chaves2022}).

With a slight abuse of notation, we use the notation of  $Y_{i,j}$ and $\hat{Y}_{i,j}$ 
that was used in the training. Whereas, in the training, the loss function was computed per training batch. Here, for accuracy evaluation, we compute the different evaluation metrics per test set. As such, let $Y_{i,j}$ and $\hat{Y}_{i,j}$ be the probability of having $j$ customers in the system in the $i^{th}$ sample of the test set of the ground truth and the ML prediction, respectively. That is, the $i^{th}$ index refers to the sample of the entire test set instead of the underlying training batch as done in Equation~\eqref{eq:loss}. For a compact vector representation let, $Y_i = (Y_{i,0}, Y_{i,1},...,Y_{i,l-1})$ and $\hat{Y}_i = (\hat{Y}_{i,0}, \hat{Y}_{i,1},...,\hat{Y}_{i,l-1})$.

 

Our first metric is the Sum of Absolute Errors (SAE) between the predicted value and the labels. It is given by\begin{align}\label{eq:SAE}
  SAE =   \frac{1}{N}\sum_{i=1}^{N}\sum_{j=0}^{l-1}|Y_{i,j}-\hat{Y}_{i,j}|,
\end{align} where $N$ represents the size of the dataset. The $SAE$ is equivalent to the Wasserstein-1 measure and (for $B=N$) corresponds to the first term of the loss function in Equation~\eqref{eq:loss}. This metric offers several advantages: (a) it is highly sensitive to any discrepancies between actual and predicted accuracy values, and (b) it yields a single value, simplifying model comparisons and making it particularly useful for tuning the model’s hyperparameters. However, since this metric provides an absolute rather than a relative value, interpreting it in the context of model accuracy evaluation can be more challenging.

The second metric, which we name $PARE$, Percentile Absolute Relative Error, measures the average relative error for a specified percentile as follows:   
\begin{align}\label{eq:InvCdf}
PARE(Y,\hat{Y}, percentile) = 100\frac{1}{N}\sum_{i=1}^{N} |\frac{F_{Y_i}^{-1}(percentile)-F_{\hat{{Y_i}}}^{-1}(percentile)}{F_{Y_i}^{-1}(percentile)}|,
\end{align} 
where $N$ represents the size of the test set and $F(\cdot)$ denotes the CDF of $Y_i$. This metric effectively identifies differences at any given percentile, including the tails of the two distributions. In our evaluations, we considered the following six percentile values: $25\%, 50\%, 75\%, 90\%, 99\%, 99.9\%$. As mentioned earlier, precise estimates of tail probabilities are crucial in many queueing applications, as they are directly related to service level measures of the system.  

Our third metric, dubbed  REM, is the Relative Error of the Mean number of customers in the system. 
\begin{align} \label{eq:REM}
REM = 100\frac{1}{N}\sum_{i=1}^N \frac{|\sum_{j=0}^{l-1}j(Y_{i,j}-\hat{Y}_{i,j})|}{\sum_{j=0}^{l-1}j\hat{Y}_{ij}}.
\end{align} 
The benefit of this measure is that it encapsulates the model's accuracy concerning the most commonly used performance metric in queueing systems—the average number of customers in the queue. Furthermore, it enables a comparison between our ML model and one of the standard approximations for $GI/GI/c$ queues.

\subsection{Moments analysis}\label{sec:mom_anal}

In this section, we optimize the value of the number of inter-arrival and service time moments, denoted as $n$. Theoretically, a larger $n$ provides more information about the queueing system, which should enhance accuracy. However, as $n$ increases, the additional information becomes less significant. Moreover, increasing $n$ may be counterproductive if it introduces more noise into the system. Therefore, we evaluate the $SAE$ values on the Test set (i) for both $GI/GI/c$ and $GI/GI_i/2$ systems. We compare all possible values within the range $1 \leq n \leq 10$.

\subsection{Performence evluation}\label{sec:peva}

After training, we wish to thoroughly examine the NN model of both systems' accuracy against the Test sets (i) and (ii). Meanwhile, for the $GI/GI/c$ system, we also compare our model against the model in the literature. Specifically, we compare PARE values against Shore~\cite {doi:10.1057/jors.1988.45}, and REM values against Whitt~\cite{https://doi.org/10.1111/j.1937-5956.1993.tb00094.x}, Marchal~\cite{Marchal1985}, KLB~\cite{KLB76}, Chaves~\cite{Chaves2022} and Shore~\cite {doi:10.1057/jors.1988.45}. To summarize, Table~\ref{tab:experiments} details four experiments that we run for accuracy evaluation.

\begin{table}[!htp]\centering
\caption{Accuracy experiments}\label{tab:experiments}
\scriptsize
\begin{tabular}{|l|r|r|r|r|}\toprule
Experiment &System &Test set &Metrics \\ \hline
1 &$GI/GI/c$ &(i) &PARE, REM \\ \hline
2 &$GI/GI/c$ &(ii) &PARE, REM \\ \hline
3 & $GI/GI_i/2$ &(i) &PARE, REM \\ \hline
4 & $GI/GI_i/2$ &(ii) &PARE, REM \\ \hline
\bottomrule
\end{tabular}
\end{table}

\section{Results}\label{sec:result}

Next, we present the results of our experiments. First, we present the moment analysis in Section~\ref{sec:mom_anal_res}. Then, in Section~\ref{sec:acc_res}, we present the result of all four experiments described in Section~\ref{sec:peva}.

\subsection{Moment analysis}\label{sec:mom_anal_res}

\begin{figure}
\centering
\includegraphics[scale=0.4]{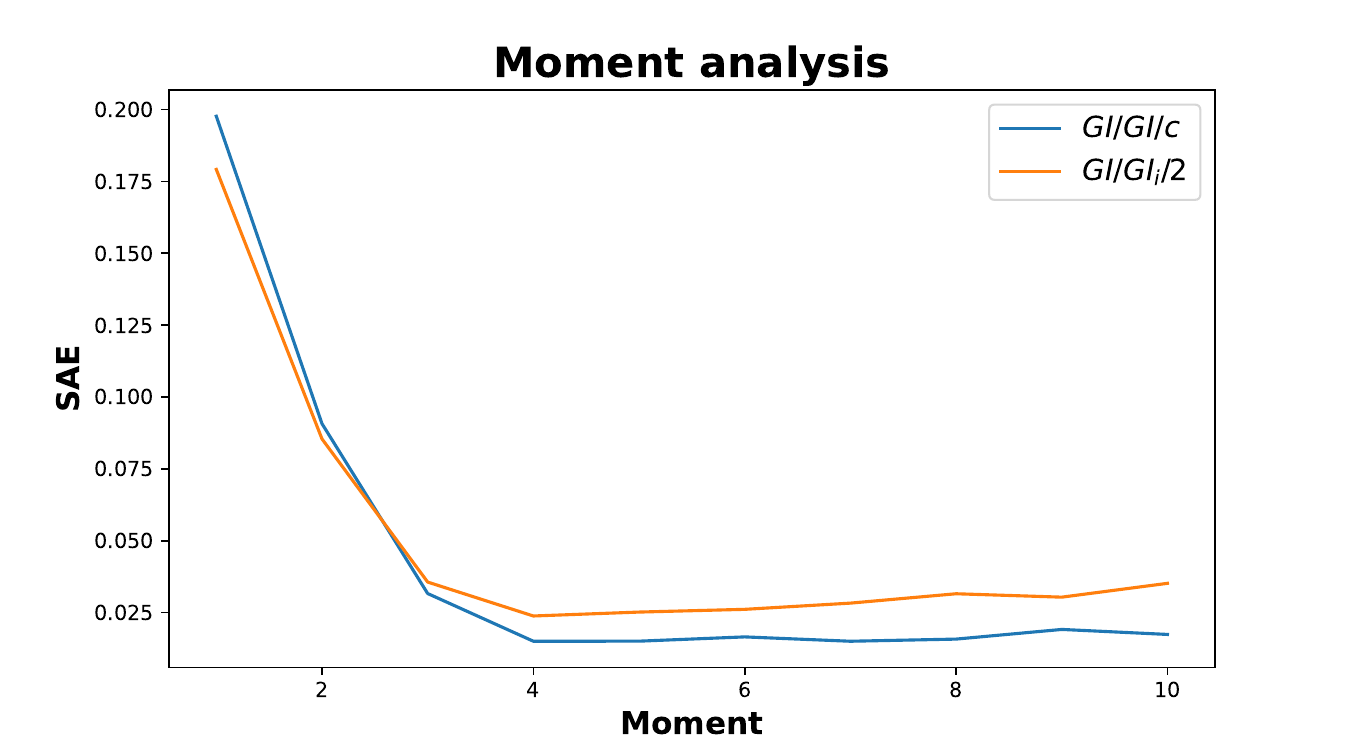}
\caption{SAE values as a function of the number of moments used as input. }
\label{fig:mom_anal}
\end{figure}

The results shown in Figure~\ref{fig:mom_anal} illustrate the accuracy, measured via the SAE metric, as a function of the number of inter-arrival and service time moments $n$ used. The figure highlights the impact of each moment on the accuracy of the $GI/GI/c$ and $GI/GI_i/2$ systems. As observed, using more than four moments introduces more noise than additional value. This is somewhat unexpected, as a similar analysis conducted for a $GI/GI/1$ system in~\cite{sherzer23} demonstrated that the fifth moment also contributed to accuracy. We hypothesize that since the labeling in~\cite{sherzer23} was performed numerically rather than through simulation, the neural network could extract useful information from the fifth moment.

Moreover, the SAE measure is not monotonic for $n \geq 5$. We believe this occurs because the inclusion of less significant data as input to the model introduces more noise than useful information.

\subsection{Accuracy results}\label{sec:acc_res}

In this Section, we present the results of our four experiments details in Section~\ref{sec:peva}.  

\subsubsection{Experiment 1}

We next show the results of Test set (i) under the $GI/GI/c$ system. In Tables~\ref{tab:PARE_low_test_1_ggc} and~\ref{tab:PARE_high_test_1_ggc}, we show the PARE results for $\{25,50,75\}$, and $\{90,99,99.9\}$ percentiles, respecively.  According to the segmentation detailed in Section~\ref{sec:segmentation}, we present all 32 settings. Recall that the SCV values of the inter-arrival and the service times are partitioned for $[0,1]$  and $[1,20]$, which we present as Low and High. We denote the quantile of $\rho$ as $Q(\rho)$, where 1 refers to $[0.01,0.25]$, 2 to $[0.25,0.5]$, 3 to $[0.5,0.75]$ and 4 to $[0.75,0.95]$. If the number of servers $c$ is five or less, we represent it as Low and High for six or more. We present the PARE values of our model, presented as NN, and the results from~\cite{doi:10.1057/jors.1988.45}, presented as Shore. The values that are colored in red represent the lower value between NN and Shore. If they are both 0.0, then they are colored in blue. 
 
As the result indicates, our model (NN) outperforms~\cite{doi:10.1057/jors.1988.45} (Shore), where only in three settings out of 192  (32 groups for six percentiles) the NN performed worse. The NN PARE increases with the percentiles, which makes sense, as the tail is more complex, with the highest PARE error of 10.57\%. Still, in most scenarios, the PARE error is lower than 2\%. The results do not indicate a significant difference between groups of SCV values or utilization levels. 


\begin{table}[!htp]\centering
\caption{$GI/GI/c$: PARE Test set (i) for percentiles: $\{25,50,75\}$ }\label{tab:PARE_low_test_1_ggc}
\scriptsize
\begin{tabular}{|l|r|r|r|r|rr|rr|rr|r|}\toprule
\textbf{} &\multicolumn{2}{|c|}{\textbf{SCV}} &\textbf{} &\textbf{} &\multicolumn{2}{|c|}{\textbf{25}} &\multicolumn{2}{|c|}{\textbf{50}} &\multicolumn{2}{|c|}{\textbf{75}} \\ \hline
\textbf{\#} &\textbf{Arrival} &\textbf{Service} &\textbf{$Q(\rho)$ } &\textbf{$c$} &NN &Shore &NN &Shore &NN &Shore \\
\midrule
1 &Low &Low &1 &Low &\textcolor{blue}{0.00} &\textcolor{blue}{0.00} &\textcolor{red}{0.00} &5.26 &\textcolor{blue}{0.00} &\textcolor{blue}{0.00} \\
2 &Low &Low &1 &High &\textcolor{red}{0.00} &7.72 &\textcolor{red}{0.00} &3.66 &\textcolor{red}{0.00} &4.20 \\
3 &Low &Low &2 &Low &\textcolor{red}{0.00} &11.68 &\textcolor{red}{0.00} &3.16 &\textcolor{red}{0.51} &5.84 \\
4 &Low &Low &2 &High &\textcolor{red}{0.00} &9.29 &\textcolor{red}{0.00} &0.59 &\textcolor{red}{0.00} &3.76 \\
5 &Low &Low &3 &Low &\textcolor{red}{0.00} &4.89 &\textcolor{red}{0.00} &3.59 &\textcolor{red}{1.70} &6.12 \\
6 &Low &Low &3 &High &\textcolor{red}{0.00} &8.07 &\textcolor{red}{0.00} &1.16 &\textcolor{red}{0.00} &8.72 \\
7 &Low &Low &4 &Low &\textcolor{red}{0.88} &6.10 &\textcolor{red}{0.75} &21.21 &\textcolor{red}{1.36} &43.10 \\
8 &Low &Low &4 &High &\textcolor{red}{0.00} &4.89 &\textcolor{red}{1.13} &11.77 &\textcolor{red}{0.98} &20.35 \\
9 &Low &High &1 &Low &\textcolor{blue}{0.00} &\textcolor{blue}{0.00} &\textcolor{red}{0.00} &4.17 &0.27 &\textcolor{red}{0.00} \\
10 &Low &High &1 &High &\textcolor{red}{0.00} &5.79 &\textcolor{red}{0.00} &2.11 &\textcolor{blue}{0.00} &\textcolor{blue}{0.00} \\
11 &Low &High &2 &Low &\textcolor{red}{0.00} &2.36 &\textcolor{red}{0.23} &3.30 &\textcolor{red}{0.43} &1.57 \\
12 &Low &High &2 &High &\textcolor{red}{0.00} &3.58 &0.34 &\textcolor{red}{0.00} &0.23 &\textcolor{red}{0.00} \\
13 &Low &High &3 &Low &\textcolor{red}{0.64} &2.43 &\textcolor{red}{0.76} &3.13 &\textcolor{red}{1.27} &21.38 \\
14 &Low &High &3 &High &\textcolor{red}{0.92} &2.94 &\textcolor{red}{0.46} &0.36 &\textcolor{red}{1.22} &3.61 \\
15 &Low &High &4 &Low &\textcolor{red}{0.51} &5.76 &\textcolor{red}{1.95} &41.85 &\textcolor{red}{1.09} &73.07 \\
16 &Low &High &4 &High &\textcolor{red}{0.00} &1.92 &\textcolor{red}{1.21} &13.53 &\textcolor{red}{1.70} &40.66 \\
17 &High &Low &1 &Low &\textcolor{blue}{0.00} &\textcolor{blue}{0.00} &\textcolor{red}{0.00} &6.77 &\textcolor{red}{0.00} &4.69 \\
18 &High &Low &1 &High &\textcolor{red}{0.00} &23.49 &\textcolor{red}{0.54} &14.46 &\textcolor{red}{1.29} &4.17 \\
19 &High &Low &2 &Low &\textcolor{red}{0.00} &20.67 &\textcolor{red}{0.82} &14.53 &\textcolor{red}{1.93} &10.42 \\
20 &High &Low &2 &High &\textcolor{red}{1.33} &52.33 &\textcolor{red}{2.52} &6.00 &\textcolor{red}{0.71} &11.49 \\
21 &High &Low &3 &Low &\textcolor{red}{2.02} &19.27 &\textcolor{red}{0.73} &12.36 &\textcolor{red}{1.25} &42.71 \\
22 &High &Low &3 &High &\textcolor{red}{0.98} &23.05 &\textcolor{red}{0.79} &7.39 &\textcolor{red}{1.25} &20.64 \\
23 &High &Low &4 &Low &\textcolor{red}{1.55} &25.17 &\textcolor{red}{1.15} &58.96 &\textcolor{red}{0.65} &78.70 \\
24 &High &Low &4 &High &\textcolor{red}{0.53} &11.02 &\textcolor{red}{0.68} &34.26 &\textcolor{red}{1.62} &57.03 \\
25 &High &High &1 &Low &\textcolor{blue}{0.00} &\textcolor{blue}{0.00} &\textcolor{red}{0.13} &3.45 &\textcolor{red}{0.58} &1.72 \\
26 &High &High &1 &High &\textcolor{red}{0.71} &12.78 &\textcolor{red}{0.65} &11.39 &\textcolor{red}{0.41} &4.86 \\
27 &High &High &2 &Low &\textcolor{red}{0.13} &12.28 &\textcolor{red}{0.86} &7.02 &\textcolor{red}{1.36} &9.26 \\
28 &High &High &2 &High &\textcolor{red}{0.44} &22.50 &\textcolor{red}{0.53} &7.13 &\textcolor{red}{0.73} &7.17 \\
29 &High &High &3 &Low &\textcolor{red}{0.15} &15.79 &\textcolor{red}{0.87} &11.91 &\textcolor{red}{1.45} &44.06 \\
30 &High &High &3 &High &\textcolor{red}{0.43} &14.15 &\textcolor{red}{0.47} &1.47 &\textcolor{red}{1.46} &16.25 \\
31 &High &High &4 &Low &\textcolor{red}{1.83} &19.69 &\textcolor{red}{0.86} &62.27 &\textcolor{red}{1.26} &81.44 \\
32 &High &High &4 &High &\textcolor{red}{0.76} &7.57 &\textcolor{red}{1.93} &27.28 &\textcolor{red}{0.75} &60.45 \\
\bottomrule
\end{tabular}
\end{table}


\begin{table}[!htp]\centering
\caption{$GI/GI/c$: PARE Test set (i) for percentiles: $\{90,99,99.9\}$}\label{tab:PARE_high_test_1_ggc}
\scriptsize
\begin{tabular}{|l|r|r|r|r|rr|rr|rr|r|}\toprule
\textbf{} &\multicolumn{2}{c}{\textbf{SCV}} &\textbf{} &\textbf{} &\multicolumn{2}{|c|}{\textbf{90}} &\multicolumn{2}{|c|}{\textbf{99}} &\multicolumn{2}{|c|}{\textbf{99.9}} \\ \hline
\textbf{\#} &\textbf{Arrival} &\textbf{Service} &\textbf{$Q(\rho)$} &\textbf{$c$} &NN &Shore &NN &Shore &NN &Shore \\ \hline
1 &Low &Low &1 &Low &\textcolor{red}{0.00} &4.21 &\textcolor{red}{1.70} &25.44 &\textcolor{red}{7.03} &23.23 \\
2 &Low &Low &1 &High &\textcolor{red}{1.75} &9.55 &\textcolor{red}{2.81} &19.07 &\textcolor{red}{2.28} &39.84 \\
3 &Low &Low &2 &Low &\textcolor{red}{0.51} &13.69 &\textcolor{red}{1.13} &10.83 &\textcolor{red}{4.83} &12.16 \\
4 &Low &Low &2 &High &\textcolor{red}{0.00} & 15.64 &\textcolor{red}{2.91} &17.78 &\textcolor{red}{2.22} &15.08 \\
5 &Low &Low &3 &Low &\textcolor{red}{1.38} &13.57 &\textcolor{red}{2.31} &20.78 &\textcolor{red}{3.08} &24.24 \\
6 &Low &Low &3 &High &\textcolor{red}{2.03} &9.47 &\textcolor{red}{1.55} &7.99 &\textcolor{red}{5.28} &15.32 \\
7 &Low &Low &4 &Low &\textcolor{red}{1.68} &56.78 &\textcolor{red}{2.57} &66.25 &\textcolor{red}{8.72} &71.69 \\
8 &Low &Low &4 &High &\textcolor{red}{0.49} &34.91 &\textcolor{red}{0.75} &48.41 &\textcolor{red}{5.14} &58.62 \\
9 &Low &High &1 &Low &\textcolor{blue}{0.00} &\textcolor{blue}{0.00} &\textcolor{red}{1.08} &10.29 &\textcolor{red}{4.67} &10.94 \\
10 &Low &High &1 &High &\textcolor{red}{0.39} &5.00 &\textcolor{red}{1.14} &10.46 &\textcolor{red}{2.91} &20.53 \\
11 &Low &High &2 &Low &\textcolor{red}{1.62} &10.13 &\textcolor{red}{1.91} &21.85 &\textcolor{red}{2.87} &32.75 \\
12 &Low &High &2 &High &\textcolor{red}{0.83} &8.51 &\textcolor{red}{1.68} &9.64 &\textcolor{red}{3.02} &11.07 \\
13 &Low &High &3 &Low &\textcolor{red}{1.04} &36.14 &\textcolor{red}{1.32} &59.40 &\textcolor{red}{2.47} &68.49 \\
14 &Low &High &3 &High &\textcolor{red}{1.03} &11.59 &\textcolor{red}{1.97} &29.39 &\textcolor{red}{3.19} &46.61 \\
15 &Low &High &4 &Low &\textcolor{red}{0.85} &81.26 &\textcolor{red}{1.19} &89.00 &\textcolor{red}{3.66} &91.54 \\
16 &Low &High &4 &High &\textcolor{red}{0.97} &60.24 &\textcolor{red}{1.24} &79.40 &\textcolor{red}{2.79} &84.68 \\
17 &High &Low &1 &Low &\textcolor{red}{2.97} &4.99 &\textcolor{red}{7.78} &16.55 &\textcolor{red}{10.57} &21.49 \\
18 &High &Low &1 &High &\textcolor{red}{0.36} &9.82 &\textcolor{red}{3.69} &17.40 &\textcolor{red}{8.06} &19.72 \\
19 &High &Low &2 &Low &\textcolor{red}{2.24} &22.31 &\textcolor{red}{3.75} &31.75 &\textcolor{red}{4.84} &33.40 \\
20 &High &Low &2 &High &\textcolor{red}{1.78} &16.69 &\textcolor{red}{2.55} &25.26 &\textcolor{red}{3.83} &28.58 \\
21 &High &Low &3 &Low &\textcolor{red}{1.64} &52.51 &\textcolor{red}{2.23} &62.39 &\textcolor{red}{2.98} &66.03 \\
22 &High &Low &3 &High &\textcolor{red}{1.26} &32.85 &\textcolor{red}{1.27} &49.59 &\textcolor{red}{2.31} &56.21 \\
23 &High &Low &4 &Low &\textcolor{red}{0.82} &83.02 &\textcolor{red}{1.32} &88.21 &\textcolor{red}{3.83} &90.01 \\
24 &High &Low &4 &High &\textcolor{red}{1.70} &68.35 &\textcolor{red}{1.99} &80.56 &\textcolor{red}{4.93} &84.45 \\
25 &High &High &1 &Low &\textcolor{red}{1.39} &5.32 &\textcolor{red}{4.01} &11.84 &\textcolor{red}{5.39} &20.37 \\
26 &High &High &1 &High &\textcolor{red}{1.12} &4.92 &\textcolor{red}{2.97} &12.40 &\textcolor{red}{5.74} &16.62 \\
27 &High &High &2 &Low &\textcolor{red}{2.12} &24.17 &\textcolor{red}{1.72} &41.29 &\textcolor{red}{2.17} &49.28 \\
28 &High &High &2 &High &\textcolor{red}{1.57} &9.73 &\textcolor{red}{2.81} &24.64 &\textcolor{red}{2.91} &33.39 \\
29 &High &High &3 &Low &\textcolor{red}{0.90} &60.23 &\textcolor{red}{1.08} &72.11 &\textcolor{red}{1.93} &76.40 \\
30 &High &High &3 &High &\textcolor{red}{1.62} &33.86 &\textcolor{red}{1.42} &57.58 &\textcolor{red}{2.31} &66.40 \\
31 &High &High &4 &Low &\textcolor{red}{1.02} &87.45 &\textcolor{red}{1.49} &91.47 &\textcolor{red}{2.96} &92.76 \\
32 &High &High &4 &High &\textcolor{red}{1.02} &74.92 &\textcolor{red}{1.27} &85.21 &\textcolor{red}{2.48} &88.83 \\
\bottomrule
\end{tabular}
\end{table}

In Table~\ref{tab:REM_test_1_GGc}, the REM values for Test set (i) are presented. To facilitate comparison, we highlight in red the model with the lowest REM. As observed, our model achieved the lowest REM in 30 out of 32 cases, ranking second in the remaining two. While traditional models occasionally exhibit low REM values, only the NN model consistently maintains low REM across all cases. As before, there are no significant differences between different SCV, $\rho$, and the number of server groups.


\begin{table}[!htp]\centering
\caption{$GI/GI/c$: REM Test set (i)}\label{tab:REM_test_1_GGc}
\scriptsize
\begin{tabular}{|l|r|r|r|r|rrrrrr|r|}\toprule
&\multicolumn{2}{|c|}{SCV} & & &\multicolumn{6}{|c|}{REM} \\ \hline
&Arrive &Service &$Q(\rho)$ &$c$ &Whitt &March &KLB &Chaves &Shore &NN \\ \hline
1 &Low &Low &1 &Low &77.4 &65.5 &60.9 &74.2 &3.9 &\textcolor{red}{1.95} \\
2 &Low &Low &1 &High &75.1 &86.4 &90.6 &96.4 &\textcolor{red}{0.1} &0.46 \\
3 &Low &Low &2 &Low &45 &25.8 &32.3 &46 &20.4 &\textcolor{red}{1.02} \\
4 &Low &Low &2 &High &59.5 &58.6 &68.5 &89.3 &2.4 &\textcolor{red}{0.55} \\
5 &Low &Low &3 &Low &23.8 &15.1 &24 &25.3 &46.2 &\textcolor{red}{1.06} \\
6 &Low &Low &3 &High &46.7 &34.5 &43.2 &72.9 &10.1 &\textcolor{red}{0.5} \\
7 &Low &Low &4 &Low &10.1 &4.5 &7.6 &25.8 &74.6 &\textcolor{red}{1.82} \\
8 &Low &Low &4 &High &11.7 &3.1 &6.1 &34.5 &50.8 &\textcolor{red}{0.8} \\
9 &Low &High &1 &Low &67 &47.8 &46.3 &49.9 &4.6 &\textcolor{red}{0.83} \\
10 &Low &High &1 &High &83.9 &68.8 &74.6 &82.6 &\textcolor{red}{0.1} &0.29 \\
11 &Low &High &2 &Low &31.5 &24.2 &22.4 &26.9 &26 &\textcolor{red}{0.86} \\
12 &Low &High &2 &High &71.8 &50.8 &47.1 &55.7 &2.2 &\textcolor{red}{0.3} \\
13 &Low &High &3 &Low &14.5 &8.5 &7.9 &17.8 &63.1 &\textcolor{red}{0.95} \\
14 &Low &High &3 &High &38.4 &28.1 &27.4 &40.6 &24.7 &\textcolor{red}{0.74} \\
15 &Low &High &4 &Low &6.6 &2.8 &3.1 &11.9 &85.7 &\textcolor{red}{1.01} \\
16 &Low &High &4 &High &15.8 &10.5 &10.6 &18.7 &56.5 &\textcolor{red}{1.17} \\
17 &High &Low &1 &Low &73.2 &58.5 &67.2 &45.5 &17.5 &\textcolor{red}{4.87} \\
18 &High &Low &1 &High &83.6 &90.7 &95.5 &77.8 &3 &\textcolor{red}{1.09} \\
19 &High &Low &2 &Low &37.3 &32 &57.3 &31 &43.3 &\textcolor{red}{2.85} \\
20 &High &Low &2 &High &64.2 &74 &87.7 &47.3 &20.3 &\textcolor{red}{1.09} \\
21 &High &Low &3 &Low &27 &26.1 &38.4 &32.1 &70.6 &\textcolor{red}{1.56} \\
22 &High &Low &3 &High &32.4 &41.2 &61.8 &48.3 &45.5 &\textcolor{red}{1.02} \\
23 &High &Low &4 &Low &8.9 &10.4 &19.2 &33.2 &87 &\textcolor{red}{0.89} \\
24 &High &Low &4 &High &16.2 &21.5 &34.1 &59.5 &67.9 &\textcolor{red}{1.45} \\
25 &High &High &1 &Low &67.1 &55.4 &58.4 &117.7 &18.5 &\textcolor{red}{2.5} \\
26 &High &High &1 &High &77.5 &79.5 &88 &45.6 &1.8 &\textcolor{red}{0.74} \\
27 &High &High &2 &Low &39.8 &42.1 &39.6 &67.3 &48.5 &\textcolor{red}{1.52} \\
28 &High &High &2 &High &47.7 &47.2 &66.5 &28.9 &14.9 &\textcolor{red}{0.88} \\
29 &High &High &3 &Low &22.5 &21.7 &21.9 &17.7 &73.9 &\textcolor{red}{0.98} \\
30 &High &High &3 &High &18.8 &19 &38.9 &17.6 &44.8 &\textcolor{red}{0.93} \\
31 &High &High &4 &Low &13.4 &11 &11.7 &13.3 &87.6 &\textcolor{red}{1.13} \\
32 &High &High &4 &High &9 &8.1 &16.2 &17.5 &71.2 &\textcolor{red}{1.06} \\
\bottomrule
\end{tabular}
\end{table}

\subsubsection{Experiment 2}

We next show the results of Test set (ii) under the $GI/GI/c$ system. In Tables~\ref{tab:PARE_low_test_2_ggc} and~\ref{tab:PARE_high_test_2_ggc}, we show the PARE results for $\{25,50,75\}$, and $\{90,99,99.9\}$ percentiles, respecively.  According to the segmentation detailed in Section~\ref{sec:segmentation}, we present all 32 settings. The column's names are as in Experiment 1. Again, the values that are colored in red represent the lower value between NN and Shore. If they are both 0.0, then they are colored in blue. 
 
As the results indicate, our model (NN) outperforms~\cite{doi:10.1057/jors.1988.45} (Shore), where the NN performed worse in only four out of 192 cases. Similarly to Test set (i), the NN PARE decreases with the percentiles, with the highest PARE error at 8.69\%. The PARE values in Test Set (ii) are, in general, lower than in Test Set (i), perhaps because the inter-arrival and service times distributions come with a simpler shape.  Again, the results do not indicate a significant difference between groups of SCV values or utilization levels.

\begin{table}[!htp]\centering
\caption{$GI/GI/c$: PARE Test set (ii) for percentiles: $\{25,50,75\}$}\label{tab:PARE_low_test_2_ggc}
\scriptsize
\begin{tabular}{|l|r|r|r|r|rr|rr|rr|r}\toprule
&\multicolumn{2}{|c|}{SCV } & & &\multicolumn{2}{|c|}{\textbf{25}} &\multicolumn{2}{|c|}{\textbf{50}} &\multicolumn{2}{|c|}{\textbf{75}} \\ \hline
\# &Arrive &Service &$Q(\rho)$ &Ser &NN &Shore &NN &Shore &NN &Shore \\
\midrule
1 &Low &Low &1 &Low &\textcolor{blue}{0.00} &\textcolor{blue}{0.00} &\textcolor{red}{0.71} &5.26 &\textcolor{blue}{0.00} &\textcolor{blue}{0.00} \\
2 &Low &Low &1 &High &\textcolor{red}{0.00} &7.72 & \textcolor{red}{1.42} &3.66 &\textcolor{red}{0.00} &4.20 \\
3 &Low &Low &2 &Low &\textcolor{red}{0.00} &11.68 &\textcolor{red}{0.55} &3.16 &\textcolor{red}{0.73} &5.84 \\
4 &Low &Low &2 &High &\textcolor{red}{1.77} &9.29 &0.83 &\textcolor{red}{0.59} &\textcolor{red}{1.77} &3.76 \\
5 &Low &Low &3 &Low &\textcolor{red}{0.83} &4.89 &\textcolor{red}{0.25} &3.59 &\textcolor{red}{0.74} &6.12 \\
6 &Low &Low &3 &High & \textcolor{red}{1.47} &8.07 &\textcolor{red}{0.92} &1.16 &\textcolor{red}{0.72} &8.72 \\
7 &Low &Low &4 &Low &\textcolor{red}{0.23} &6.10 &\textcolor{red}{1.95} &21.21 &\textcolor{red}{1.91} &43.10 \\
8 &Low &Low &4 &High &\textcolor{red}{0.43} &4.89 &\textcolor{red}{0.62} &11.77 &\textcolor{red}{1.45} &20.35 \\
9 &Low &High &1 &Low &\textcolor{blue}{0.00} &\textcolor{blue}{0.00} &\textcolor{red}{0.49} &4.17 &\textcolor{blue}{0.00} &\textcolor{blue}{0.00} \\
10 &Low &High &1 &High &\textcolor{blue}{0.00} &5.79 &\textcolor{red}{0.00} &2.11 &\textcolor{blue}{0.00} &\textcolor{blue}{0.00} \\
11 &Low &High &2 &Low &\textcolor{blue}{0.00} &2.36 &\textcolor{blue}{0.00} &3.30 &\textcolor{red}{0.44} &1.57 \\
12 &Low &High &2 &High &\textcolor{red}{0.22} &3.58 &0.22 &\textcolor{red}{0.00} &0.88 &\textcolor{red}{0.00} \\
13 &Low &High &3 &Low & \textcolor{red}{2.08} &2.43 &\textcolor{blue}{0.00} &3.13 &\textcolor{blue}{0.00} &21.38 \\
14 &Low &High &3 &High &\textcolor{red}{0.44} &2.94 &\textcolor{red}{0.16} &0.36 &\textcolor{red}{0.46} &3.61 \\
15 &Low &High &4 &Low &\textcolor{red}{0.96} &5.76 &\textcolor{red}{1.32} &41.85 &\textcolor{red}{2.26} &73.07 \\
16 &Low &High &4 &High &\textcolor{red}{0.35} &1.92 &\textcolor{red}{0.68} &13.53 &\textcolor{red}{0.86} &40.66 \\
17 &High &Low &1 &Low &\textcolor{blue}{0.00} &\textcolor{blue}{0.00} &\textcolor{red}{0.00} &6.77 &\textcolor{blue}{0.00} &4.69 \\
18 &High &Low &1 &High &\textcolor{red}{0.00} &23.49 &\textcolor{red}{1.50} &14.46 &\textcolor{red}{0.38} &4.17 \\
19 &High &Low &2 &Low &\textcolor{red}{0.70} &20.67 &\textcolor{red}{0.00} &14.53 &\textcolor{red}{0.18} &10.42 \\
20 &High &Low &2 &High &\textcolor{red}{0.33} &52.33 &\textcolor{red}{1.34} &6.00 &\textcolor{red}{0.59} &11.49 \\
21 &High &Low &3 &Low &\textcolor{red}{0.38} &19.27 &\textcolor{red}{0.68} &12.36 &\textcolor{red}{0.48} &42.71 \\
22 &High &Low &3 &High &\textcolor{red}{0.47} &23.05 &\textcolor{red}{0.75} &7.39 &\textcolor{red}{1.20} &20.64 \\
23 &High &Low &4 &Low &\textcolor{red}{1.80} &25.17 &\textcolor{red}{2.16} &58.96 &\textcolor{red}{2.12} &78.70 \\
24 &High &Low &4 &High &\textcolor{red}{0.94} &11.02 &\textcolor{red}{1.04} &34.26 &\textcolor{red}{0.99} &57.03 \\
25 &High &High &1 &Low &\textcolor{blue}{0.00} &\textcolor{blue}{0.00} &\textcolor{red}{0.00} &3.45 &\textcolor{red}{0.00} &1.72 \\
26 &High &High &1 &High &\textcolor{red}{0.34} &12.78 &\textcolor{red}{0.89} &11.39 &\textcolor{red}{2.35} &4.86 \\
27 &High &High &2 &Low &\textcolor{red}{0.00} &12.28 &\textcolor{red}{0.42} &7.02 &\textcolor{red}{0.00} &9.26 \\
28 &High &High &2 &High &\textcolor{red}{0.59} &22.50 &\textcolor{red}{0.67} &7.13 &\textcolor{red}{0.14} &7.17 \\
29 &High &High &3 &Low &\textcolor{red}{1.18} &15.79 &\textcolor{red}{0.81} &11.91 &\textcolor{red}{1.54} &44.06 \\
30 &High &High &3 &High &\textcolor{red}{0.43} &14.15 &\textcolor{red}{0.70} &1.47 &\textcolor{red}{1.38} &16.25 \\
31 &High &High &4 &Low &\textcolor{red}{1.82} &19.69 &\textcolor{red}{1.34} &62.27 &\textcolor{red}{1.73} &81.44 \\
32 &High &High &4 &High &\textcolor{red}{0.15} &7.57 &\textcolor{red}{1.31} &27.28 &\textcolor{red}{1.64} &60.45 \\
\bottomrule
\end{tabular}
\end{table}

\begin{table}[!htp]\centering
\caption{$GI/GI/c$: PARE Test set (ii) for percentiles: $\{90,99,99.9\}$}\label{tab:PARE_high_test_2_ggc}
\scriptsize
\begin{tabular}{|l|r|r|r|r|rr|rr|rr|r}\toprule
&\multicolumn{2}{|c|}{SCV } & & &\multicolumn{2}{|c|}{\textbf{90}} &\multicolumn{2}{|c|}{\textbf{99}} &\multicolumn{2}{|c|}{\textbf{99.9}} \\ \hline
\# &Arrive &Service &$Q(\rho)$  &$c$ &NN &Shore &NN &Shore &NN &Shore \\
\midrule
1  &Low  &Low  &1  &Low  &\textcolor{red}{0.48} &4.21  &\textcolor{red}{7.36} &25.44  &\textcolor{red}{1.24} &23.23  \\
2  &Low  &Low  &1  &High &\textcolor{red}{0.57} &9.55  &\textcolor{red}{2.39} &19.07  &\textcolor{red}{2.16} &39.84  \\
3  &Low  &Low  &2  &Low  &\textcolor{red}{0.95} &13.69 &\textcolor{red}{2.10} &10.83  &\textcolor{red}{2.73} &12.16  \\
4  &Low  &Low  &2  &High &\textcolor{red}{0.53} &15.64 &\textcolor{red}{2.07} &17.78  &\textcolor{red}{5.38} &15.08  \\
5  &Low  &Low  &3  &Low  &\textcolor{red}{0.92} &13.57 &\textcolor{red}{1.98} &20.78  &\textcolor{red}{5.28} &24.24  \\
6  &Low  &Low  &3  &High &\textcolor{red}{1.41} &9.47  &\textcolor{red}{3.31} &7.99   &\textcolor{red}{5.81} &15.32  \\
7  &Low  &Low  &4  &Low  &\textcolor{red}{3.05} &56.78 &\textcolor{red}{3.14} &66.25  &\textcolor{red}{7.96} &71.69  \\
8  &Low  &Low  &4  &High &\textcolor{red}{2.57} &34.91 &\textcolor{red}{5.02} &48.41  &\textcolor{red}{8.69} &58.62  \\
9  &Low  &High &1  &Low  &0.97 &\textcolor{red}{0.00}  &\textcolor{red}{4.94} &10.29  &\textcolor{red}{0.86} &10.94  \\
10 &Low  &High &1  &High &\textcolor{red}{0.00} &5.00  &\textcolor{red}{1.08} &10.46  &\textcolor{red}{1.51} &20.53  \\
11 &Low  &High &2  &Low  &\textcolor{red}{2.66} &10.13 &\textcolor{red}{2.42} &21.85  &\textcolor{red}{4.67} &32.75  \\
12 &Low  &High &2  &High &\textcolor{red}{0.45} &8.51  &\textcolor{red}{1.48} &9.64   &\textcolor{red}{3.78} &11.07  \\
13 &Low  &High &3  &Low  &\textcolor{red}{2.31} &36.14 &\textcolor{red}{2.20} &59.40  &\textcolor{red}{5.59} &68.49  \\
14 &Low  &High &3  &High &\textcolor{red}{0.71} &11.59 &\textcolor{red}{2.28} &29.39  &\textcolor{red}{5.10} &46.61  \\
15 &Low  &High &4  &Low  &\textcolor{red}{1.60} &81.26 &\textcolor{red}{2.35} &89.00  &\textcolor{red}{8.27} &91.54  \\
16 &Low  &High &4  &High &\textcolor{red}{2.01} &60.24 &\textcolor{red}{4.31} &79.40  &\textcolor{red}{7.20} &84.68  \\
17 &High &Low  &1  &Low  &\textcolor{red}{0.41} &4.99  &\textcolor{red}{6.78} &16.55  &\textcolor{red}{3.51} &21.49  \\
18 &High &Low  &1  &High &\textcolor{red}{2.26} &9.82  &\textcolor{red}{3.38} &17.40  &\textcolor{red}{3.34} &19.72  \\
19 &High &Low  &2  &Low  &\textcolor{red}{1.91} &22.31 &\textcolor{red}{3.89} &31.75  &\textcolor{red}{4.23} &33.40  \\
20 &High &Low  &2  &High &\textcolor{red}{1.13} &16.69 &\textcolor{red}{2.77} &25.26  &\textcolor{red}{4.13} &28.58  \\
21 &High &Low  &3  &Low  &\textcolor{red}{1.79} &52.51 &\textcolor{red}{3.04} &62.39  &\textcolor{red}{4.56} &66.03  \\
22 &High &Low  &3  &High &\textcolor{red}{1.52} &32.85 &\textcolor{red}{2.40} &49.59  &\textcolor{red}{3.79} &56.21  \\
23 &High &Low  &4  &Low  &\textcolor{red}{2.39} &83.02 &\textcolor{red}{3.38} &88.21  &\textcolor{red}{6.49} &90.01  \\
24 &High &Low  &4  &High &\textcolor{red}{1.81} &68.35 &\textcolor{red}{2.85} &80.56  &\textcolor{red}{6.42} &84.45  \\
25 &High &High &1  &Low  &\textcolor{red}{2.46} &5.32  &\textcolor{red}{7.65} &11.84  &\textcolor{red}{3.64} &20.37  \\
26 &High &High &1  &High &\textcolor{red}{1.78} &4.92  &\textcolor{red}{2.36} &12.40  &\textcolor{red}{3.23} &16.62  \\
27 &High &High &2 &Low  &\textcolor{red}{0.95} &24.17 &\textcolor{red}{3.56} &41.29  &\textcolor{red}{5.06} &49.28  \\
28 &High &High &2 &High &\textcolor{red}{1.23} &9.73  &\textcolor{red}{3.51} &24.64  &\textcolor{red}{4.00} &33.39  \\
29 &High &High &3 &Low  &\textcolor{red}{1.97} &60.23 &\textcolor{red}{2.78} &72.11  &\textcolor{red}{5.12} &76.40  \\
30 &High &High &3 &High &\textcolor{red}{1.46} &33.86 &\textcolor{red}{2.36} &57.58  &\textcolor{red}{3.70} &66.40  \\
31 &High &High &4 &Low  &\textcolor{red}{1.77} &87.45 &\textcolor{red}{3.02} &91.47  &\textcolor{red}{5.90} &92.76  \\
32 &High &High &4 &High &\textcolor{red}{1.83} &74.92 &\textcolor{red}{2.88} &85.21  &\textcolor{red}{6.12} &88.83  \\

\bottomrule
\end{tabular}
\end{table}

In Table~\ref{tab:REM_test_2_ggc}, the REM values for Test set (ii) are presented. Our model achieved the lowest REM in 26 out of 32 cases, ranking second in the remaining. The maximum error obtained by our model is 1.9\%. As before, there are no significant differences between different SCV, $\rho$, and the number of server groups.

\begin{table}[!htp]\centering
\caption{$GI/GI/c$: REM Test set (ii)}\label{tab:REM_test_2_ggc}
\scriptsize
\begin{tabular}{|l|r|r|r|r|rrrrrr|r|}\toprule
&\multicolumn{2}{c}{SCV} & & &\multicolumn{6}{c}{REM} \\ \hline
&Arrive &Ser &$ Q(\rho)$ & $c$ &Whitt &March &KLB &Chaves &Shore &NN \\
\midrule
1  &Low  &Low  &1  &Low  &95.9  &98.9  &86.8  &81.5  &0.7  &\textcolor{red}{0.3}  \\
2  &Low  &Low  &1  &High &100   &100   &100   &100   &\textcolor{red}{0.0}  &0.2  \\
3  &Low  &Low  &2  &Low  &90.6  &85.7  &65.9  &40.1  &1.1  &\textcolor{red}{0.6}  \\
4  &Low  &Low  &2  &High &99.7  &99.9  &99.9  &91.3  &\textcolor{red}{0.1}  &0.4  \\
5  &Low  &Low  &3  &Low  &74.2  &38.9  &23.1  &45.2  &10.0  &\textcolor{red}{0.8}  \\
6  &Low  &Low  &3  &High &95.5  &83.5  &81.5  &60.0  &0.9  &\textcolor{red}{0.7}  \\
7  &Low  &Low  &4  &Low  &22.6  &6.2   &4.9   &57.4  &44.8  &\textcolor{red}{1.9}  \\
8  &Low  &Low  &4  &High &28.4  &17.0  &15.5  &89.6  &20.1  & \textcolor{red}{1.7}  \\
9  &Low  &High &1  &Low  &73.7  &73.9  &78.2  &73.8  &1.9  &\textcolor{red}{0.2}  \\
10 &Low  &High &1  &High &100   &100   &100   &100   &\textcolor{red}{0.0}  &0.2  \\
11 &Low  &High &2  &Low  &57.8  &41.5  &55.5  &40.6  &9.3  &\textcolor{red}{0.6}  \\
12 &Low  &High &2  &High &98.6  &94.6  &98.8  &73.4  &\textcolor{red}{0.1}  &0.3  \\
13 &Low  &High &3  &Low  &34.7  &17.1  &29.0  &61.0  &35.0  &\textcolor{red}{0.6}  \\
14 &Low  &High &3  &High &81.7  &67.1  &75.8  &82.5  &5.4   &\textcolor{red}{0.5}  \\
15 &Low  &High &4  &Low  &8.1   &4.9   &7.7   &60.5  &76.0  &\textcolor{red}{1.4}  \\
16 &Low  &High &4  &High &29.0  &17.7  &22.8  &95.6  &44.3  &\textcolor{red}{1.4}  \\
17 &High &Low  &1  &Low  &79.2  &60.7  &61.1  &71.0  &7.3   &\textcolor{red}{0.5}  \\
18 &High &Low  &1  &High &76.4  &82.7  &91.1  &92.3  &\textcolor{red}{0.3}  &\textcolor{red}{0.3}  \\
19 &High &Low  &2  &Low  &46.9  &30.8  &41.6  &44.0  &19.5  &\textcolor{red}{0.9}  \\
20 &High &Low  &2  &High &53.1  &59.1  &76.4  &73.4  &3.5   &\textcolor{red}{0.5}  \\
21 &High &Low  &3  &Low  &25.0  &17.0  &25.1  &24.4  &47.2  & \textcolor{red}{1.1}  \\
22 &High &Low  &3  &High &27.4  &33.4  &52.4  &41.8  &18.8  &\textcolor{red}{0.7}  \\
23 &High &Low  &4  &Low  &10.1  &7.7   &11.1  &35.2  &79.3  &\textcolor{red}{2.3}  \\
24 &High &Low  &4  &High &10.0  &12.4  &21.7  &67.0  &54.5  & \textcolor{red}{1.5}  \\
25 &High &High &1  &Low  &98.8  &68.0  &47.6  &59.9  &5.5   &\textcolor{red}{0.4}  \\
26 &High &High &1  &High &87.5  &61.2  &82.8  &89.5  &\textcolor{red}{0.2}  &0.3  \\
27 &High &High &2  &Low  &51.8  &35.8  &24.6  &30.3  &24.2  &\textcolor{red}{1.2}  \\
28 &High &High &2  &High &56.7  &34.6  &55.7  &74.6  &3.1   &\textcolor{red}{0.5}  \\
29 &High &High &3  &Low  &22.6  &17.0  &13.9  &17.0  &55.2  & \textcolor{red}{1.3}  \\
30 &High &High &3  &High &27.2  &17.1  &28.1  &49.9  &22.0  &\textcolor{red}{0.9}  \\
31 &High &High &4  &Low  &8.7   &6.1   &6.2   &11.4  &84.0  & \textcolor{red}{1.8}  \\
32 &High &High &4  &High &13.3  &7.3   &9.8   &14.2  &60.0  &\textcolor{red}{1.5}  \\
\bottomrule
\end{tabular}
\end{table}

\subsubsection{Experiment 3}

In this section, we present the results for the $GI/GI_i/2$ system. As there are no existing models for comparison, we report the performance of our model alone. The SCV values of the inter-arrival times are listed under the 'arrive' column, while the SCV values for the two different service time distributions are shown under columns 1 and 2 (under 'SCV service'). In general, the REM measure shows good performance, with the largest error being 6.16\%. While the PARE values are primarily low, they are seldom high, with the maximum error of 14.18\%. Compared to the $GI/GI/c$, we identify an increase in the error, which is not surprising given that it is a more challenging task. In general, a large SCV value induces larger errors. We observe that in the tail of the distribution, we get larger errors when the utilization is high. This is demonstrated in the PARE values of the $99.9\%$, where the largest error is revised in most settings, when $Q(\rho)=4$. We also identify larger errors when the inter-arrival SCV is high.

\begin{table}[!htp]\centering
\caption{$GI/GI_i/2$: PARE and REM  Test set (i)}\label{tab:gg2_test1}
\scriptsize
\begin{tabular}{|l|r|r|r|r|r|r|r|r|r|r|r|r|}\toprule
&SCV &\multicolumn{2}{|c|}{SCV Service} & &\multicolumn{6}{|c|}{PARE} & \\ \hline 
&Arrive &1 &2 &$Q(\rho)$ &25 &50 &75 &90 &99 &99.9 &REM \\ \hline 
1 &Low &Low &Low &1 &0.00 &3.87 &5.05 &3.80 &2.58 &4.32 &3.76 \\
2 &Low &Low &Low &2 &0.00 &6.25 &2.78 &1.39 &5.93 &4.42 &3.11 \\
3 &Low &Low &Low &3 &0.00 &0.00 &4.17 &7.94 &0.00 &4.70 &2.02 \\
4 &Low &Low &Low &4 &0.00 &10.00 &0.00 &0.00 &3.93 &4.02 &1.87 \\
5 &Low &Low &High &1 &3.75 &6.28 &5.34 &4.72 &5.33 &9.50 &4.91 \\
6 &Low &Low &High &2 &0.00 &2.78 &2.38 &1.44 &3.73 &6.11 &2.42 \\
7 &Low &Low &High &3 &0.00 &0.00 &3.77 &3.05 &2.78 &6.47 &1.90 \\
8 &Low &Low &High &4 &0.00 &1.96 &0.00 &1.14 &2.62 &5.01 &1.43 \\
9 &Low &High &Low &1 &1.48 &3.23 &5.23 &4.61 &4.97 &11.37 &4.68 \\
10 &Low &High &Low &2 &4.00 &3.00 &1.70 &3.09 &3.41 &4.59 &2.55 \\
11 &Low &High &Low &3 &0.00 &1.15 &2.41 &5.17 &5.90 &8.19 &3.09 \\
12 &Low &High &Low &4 &0.00 &2.27 &0.00 &0.00 &3.14 &5.87 &1.79 \\
13 &Low &High &High &1 &2.59 &4.24 &4.28 &4.09 &4.95 &11.78 &4.36 \\
14 &Low &High &High &2 &0.85 &2.82 &3.49 &2.56 &3.21 &4.61 &2.52 \\
15 &Low &High &High &3 &0.00 &0.49 &3.19 &3.40 &4.23 &5.26 &2.63 \\
16 &Low &High &High &4 &0.39 &0.97 &0.26 &1.09 &3.35 &6.39 &1.59 \\
17 &High &Low &Low &1 &0.00 &0.00 &3.61 &3.09 &3.70 &4.83 &3.39 \\
18 &High &Low &Low &2 &4.76 &2.88 &4.85 &7.04 &10.70 &14.18 &6.16 \\
19 &High &Low &Low &3 &3.03 &0.00 &2.84 &6.28 &6.09 &7.18 &4.87 \\
20 &High &Low &Low &4 &0.00 &0.00 &1.61 &1.61 &9.18 &12.18 &4.52 \\
21 &High &Low &High &1 &3.04 &3.89 &4.32 &4.13 &4.38 &8.83 &4.32 \\
22 &High &Low &High &2 &3.93 &2.77 &3.03 &2.43 &3.06 &4.37 &2.71 \\
23 &High &Low &High &3 &0.53 &1.15 &3.26 &3.70 &4.54 &5.86 &3.56 \\
24 &High &Low &High &4 &0.00 &0.00 &1.08 &1.74 &7.08 &10.08 &4.04 \\
25 &High &High &Low &1 &5.73 &3.64 &4.11 &3.66 &4.14 &9.91 &3.91 \\
26 &High &High &Low &2 &1.75 &3.50 &2.57 &2.65 &3.13 &3.86 &2.52 \\
27 &High &High &Low &3 &0.59 &0.59 &2.80 &4.35 &5.97 &6.25 &4.21 \\
28 &High &High &Low &4 &0.00 &0.28 &0.00 &3.34 &8.40 &11.53 &4.75 \\
29 &High &High &High &1 &4.42 &3.36 &3.48 &3.07 &3.45 &10.27 &3.39 \\
30 &High &High &High &2 &1.33 &3.07 &2.37 &2.09 &2.46 &3.79 &2.09 \\
31 &High &High &High &3 &0.52 &0.59 &3.43 &3.02 &3.25 &4.04 &2.62 \\
32 &High &High &High &4 &0.08 &0.41 &0.77 &1.98 &5.59 &7.33 &3.46 \\
\bottomrule
\end{tabular}
\end{table}

\subsubsection{Experiment 4}

Here, we present the $GI/GI_i/2$ results of the test set (ii), following the exact format as Table~\ref{tab:gg2_test1}.  Compared to Experiment 3, there is no substantial change in the accuracy, where the largest REM error is 5.79\%, and the largest PARE value is 13.7\% in the $99.99^{th}$ percentile. Again, compared to the $GI/GI/c$, we identify an increase in the error. We observe the same error behavior as in Experiment 3 (i.e., larger errors with high utilization and large inter-arrival SCV values).

\begin{table}[!htp]\centering
\caption{$GI/GI_i/2$: PARE and REM  Test set (ii)}\label{tab:gg2_test2}
\scriptsize
\begin{tabular}{|l|r|r|r|r|r|r|r|r|r|r|r|r|}\toprule
&SCV &\multicolumn{2}{|c|}{SCV Service} & &\multicolumn{6}{|c|}{PARE} & \\ \hline
&Arrive &1 &2 &$Q(\rho)$ &25 &50 &75 &90 &99 &99.9 &REM \\ \hline
1 &Low &Low &Low &1 &0.00 &0.00 &0.00 &0.00 &0.67 &5.56 &1.80 \\
2 &Low &Low &Low &2 &2.59 &0.33 &0.43 &1.54 &2.14 &8.89 &1.26 \\
3 &Low &Low &Low &3 &1.10 &0.47 &1.41 &1.80 &3.30 &5.98 &1.78 \\
4 &Low &Low &Low &4 &0.00 &0.00 &1.54 &4.10 &2.11 &3.60 &2.16 \\
5 &Low &Low &High &1 &0.00 &0.00 &0.27 &0.00 &0.51 &4.20 &2.06 \\
6 &Low &Low &High &2 &1.43 &0.00 &1.07 &0.98 &3.86 &8.60 &1.50 \\
7 &Low &Low &High &3 &1.44 &0.96 &1.51 &3.08 &4.32 &11.02 &2.19 \\
8 &Low &Low &High &4 &3.85 &2.88 &5.38 &5.93 &4.41 &11.65 &4.46 \\
9 &Low &High &Low &1 &0.00 &0.00 &0.38 &0.29 &1.16 &3.88 &2.06 \\
10 &Low &High &Low &2 &0.00 &1.17 &0.88 &0.79 &2.54 &6.59 &1.30 \\
11 &Low &High &Low &3 &0.93 &0.52 &1.76 &3.17 &5.97 &9.52 &2.70 \\
12 &Low &High &Low &4 &0.00 &1.11 &4.65 &4.37 &7.95 &11.43 &4.44 \\
13 &Low &High &High &1 &0.00 &0.00 &0.24 &0.00 &0.20 &1.87 &1.61 \\
14 &Low &High &High &2 &1.55 &0.31 &0.93 &0.81 &3.51 &7.37 &1.32 \\
15 &Low &High &High &3 &1.50 &0.33 &3.23 &3.61 &5.39 &10.48 &2.83 \\
16 &Low &High &High &4 &2.01 &2.44 &3.58 &4.03 &6.29 &9.25 &4.29 \\
17 &High &Low &Low &1 &0.00 &0.00 &0.00 &0.97 &5.41 &10.76 &5.79 \\
18 &High &Low &Low &2 &0.00 &0.93 &1.01 &5.39 &10.63 &13.70 &5.61 \\
19 &High &Low &Low &3 &0.97 &0.76 &3.89 &6.80 &9.23 &12.23 &5.09 \\
20 &High &Low &Low &4 &1.97 &3.86 &5.06 &4.71 &5.76 &11.03 &5.03 \\
21 &High &Low &High &1 &0.00 &0.00 &1.20 &1.49 &4.57 &7.30 &4.25 \\
22 &High &Low &High &2 &0.00 &0.34 &0.77 &4.87 &9.37 &9.80 &5.14 \\
23 &High &Low &High &3 &1.22 &0.71 &4.57 &6.07 &8.43 &10.72 &5.67 \\
24 &High &Low &High &4 &2.03 &3.63 &4.53 &4.80 &5.55 &9.93 &4.95 \\
25 &High &High &Low &1 &0.00 &0.00 &0.00 &0.68 &4.63 &10.13 &4.80 \\
26 &High &High &Low &2 &0.00 &0.48 &1.23 &4.20 &9.71 &12.75 &4.75 \\
27 &High &High &Low &3 &1.46 &1.18 &4.64 &5.69 &7.65 &9.22 &5.22 \\
28 &High &High &Low &4 &3.13 &5.07 &5.03 &4.72 &5.97 &10.11 &4.88 \\
29 &High &High &High &1 &0.00 &0.00 &0.24 &0.40 &4.19 &6.39 &3.70 \\
30 &High &High &High &2 &0.00 &0.84 &1.29 &3.58 &6.93 &7.85 &4.39 \\
31 &High &High &High &3 &1.18 &1.11 &4.80 &5.54 &5.83 &7.52 &4.85 \\
32 &High &High &High &4 &3.47 &3.71 &4.20 &4.52 &4.87 &8.72 &4.39 \\
\bottomrule
\end{tabular}
\end{table}

\subsection{Runtimes}\label{sec:runtimes}

In this section, we measure the runtime of our NN for both systems. The runtimes representing the inference time for each NN are tested over a personal PC, with a processor Intel(R) Core(TM) i7-14650HX   2.20 GHz, 32 GB. Table~\ref{tab:runtimes} presents the inference time for each NN. One of the advantages of applying deep learning models is that it enables the inference to be conducted in parallel while keeping the runtimes without increasing them. As such, we examine 5000 instances in parallel each time to demonstrate its speed advantage over simulation. As the results indicate, for 5000 instances, all NNs take only a fraction of a second.

\begin{table}[!htp]\centering
\caption{NN Runtimes }\label{tab:runtimes}
\scriptsize
\begin{tabular}{lrrr}\toprule
System &Number of instances &Runtimes \\ \hline
$GI/GI/c$ &5000 &0.0021 [sec] \\ \hline
$GI/GI_i/2$ &5000 &0.0024 [sec] \\ \hline
\bottomrule
\end{tabular}
\end{table}

\section{Numerical exmaple}\label{sec:num_example}

In this Section, we illustrate the benefit of our approach by solving an optimization problem. Suppose you manage a factory and are required to design a homogeneous multi-server station in your factory line. You wish to minimize the average number of jobs waiting in the queue. The manager can decide how many servers and how fast, on average, they serve a single job. For simplicity, we assume the shape of the service time distribution remains the same and can be scaled according to the service rate. 

The more servers used, and the faster the servers are, the less customers wait. However, it comes with a cost. For this example, we assume that a single server cost is determined by its rate. Let $C_1(rate)$ be the cost of a single server, which is computed according to the following:

\begin{align}\label{eq:rate_cost}
    C_1(rate) = 500*(1+rate)^5
\end{align}

Let $L$ be the number of jobs in the system, including those in service. We assume that $C_2$ is a linear $E[L]$ cost. Thus, we conclude the total cost, which is determined by the service rate ($rate$) and the number of servers ($c$):

\begin{align}\label{eq:cost}
    \min_{rate, c } Cost(rate, c) = C_1(rate)*c +C_2*E[L]   
\end{align}

For this example, the feasible service rate domain is $[0.1,0.3]$. and $C_2 = 100$. To solve the problem, we take a brute-force approach. We will approximate the cost function for any combination of $c$ and $rate$. Since $rate$ is a continuous variable, we will scan the domain over small intervals of 0.01. Thus, we are left with 2000 combinations (that is, 200 rates for each number of servers). One of the advantages of our approach is a small running time, and it can do many in parallel. In this case, it took 0.002 seconds for all 2000 cases. 

\begin{figure}
\centering
\includegraphics[scale=0.55]{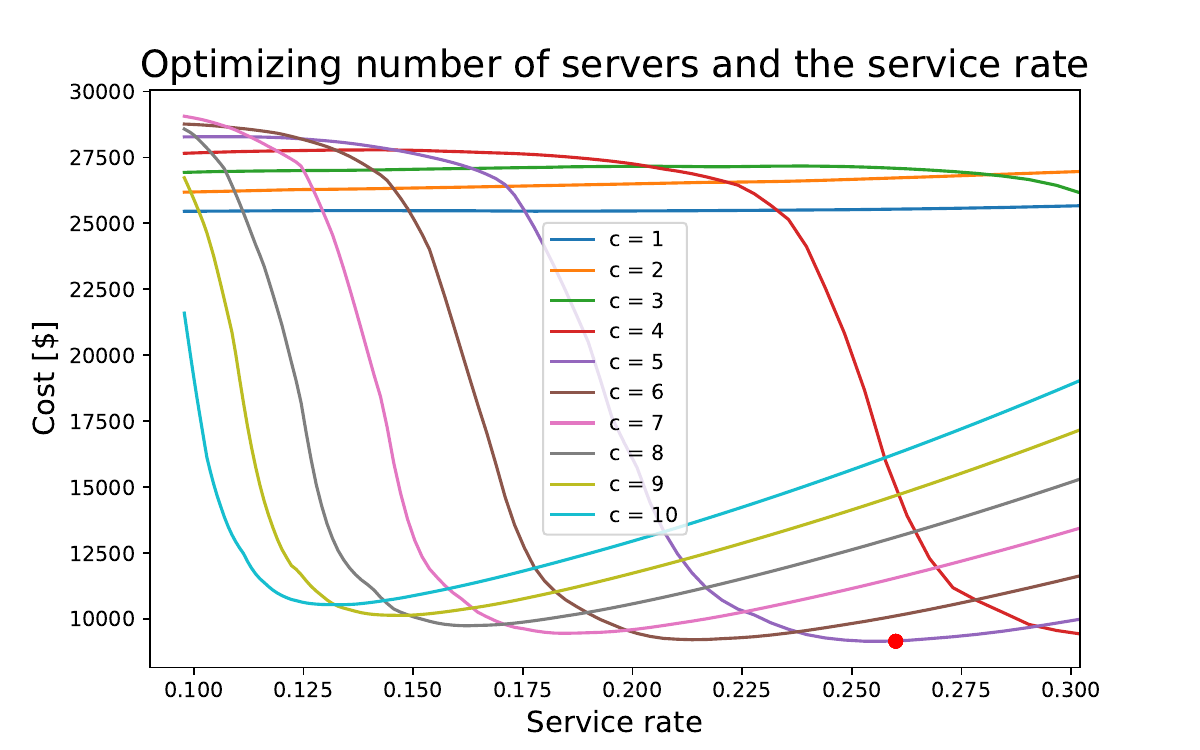}
\caption{Total cost as a function of $rate$ and $c$. }
\label{fig:opt}
\end{figure}

Figure~\ref{fig:opt} shows the cost as a function of both $rate$ and $c$. The red dot marks the minimum cost, which is 9159.6, achieved at $rate = 0.261$ and $c = 6$. This numerical example highlights the potential advantages of using machine learning for queueing system design. It also demonstrates that even a brute-force search can be practical in such problems.  

\section{Discussion}\label{sec:discussion}

Next, we discuss possible extensions in Section~\ref{sec:extensions} and limitations in Section~\ref{sec:limitations}.

\subsection{Extensions}\label{sec:extensions}
This paper is part of a series of papers that use NNs to analyze queueing systems. The complexity of this paper was the use of multi-server stations, as opposed to other papers that dealt only with single-server stations~\cite{sherzer23, sherzer2024computingsteadystateprobabilitiestandem}. Yet, this is a very important building block that can be extended to more complex queueing network topologies.

As mentioned above, in~\cite{sherzer2024computingsteadystateprobabilitiestandem}, the authors used an ML approach as here to analyze a non-Markovian tandem queueing systems. This is done by decomposing the queueing network. An NN was used for each station to approximate the departure process, which is the next station inter-arrival process, and as an input for the following station analysis. Note that the departure from any station is non-renewal. Hence, as shown in~\cite{sherzer2024computingsteadystateprobabilitiestandem}, auto-correlation values describe inter-departure dependencies.  

A straightforward extension of our approach is applying the analysis of~\cite{sherzer2024computingsteadystateprobabilitiestandem} but replace single-station queues with multi-station queues, both for homogenous and heterogeneous servers. For this, one needs to train NNs to predict the inter-departure from multi-server stations and steady-state probabilities of the number of servers in a multi-server system, as is done in this paper, only this time arrivals are according to a non-renewal process, namely a $G/GI/c$ queue. 

\begin{figure}
\centering
\includegraphics[scale=0.43]{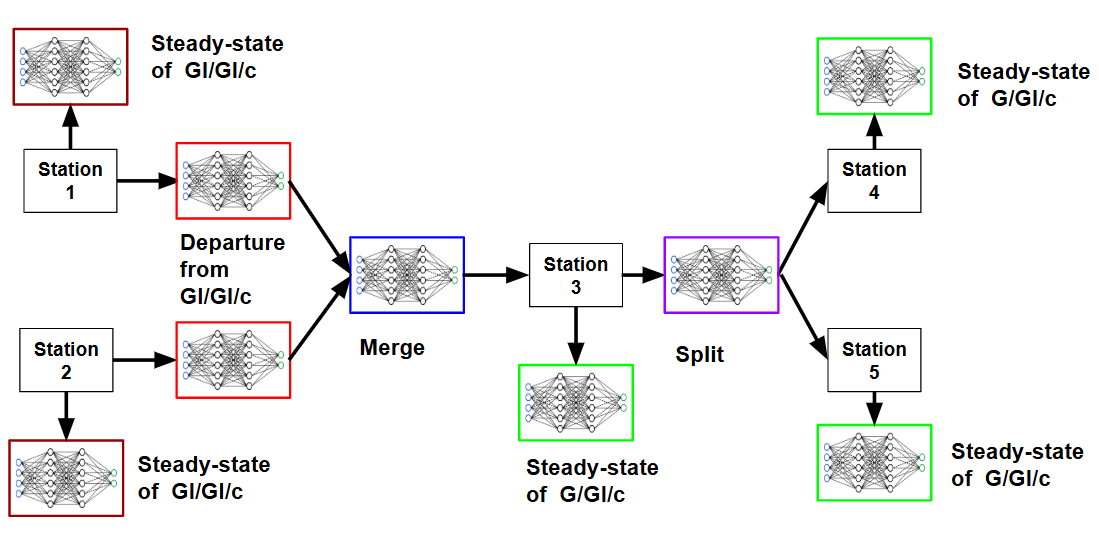}
\caption{An example of a feed-forward queueing network. }
\label{fig:feed}
\end{figure}

A more interesting extension of our method is a more general feed-forward queueing network with multiple servers. By feed-forward queueing network, we mean a network without feedback loops, meaning customers do not revisit the same station after departing. To enable this analysis, our ML approach requires two additional functionalities beyond those mentioned in a tandem system (i.e., predicting inter-departure process and steady-state for a $G/GI/c$ queue). The first is the ability to merge (superposition) two or more (possibly non-renewal) inter-arrival processes (represented by the blue NN in Figure~\ref{fig:feed}). The second is the ability to split (thinning) a departure process into at least two separate streams (represented by the purple NN in Figure~\ref{fig:feed}). In light of the results presented in this paper, applying machine learning to approximate feed-forward queueing networks with multiple servers is a feasible and promising approach.

One advantage of this decomposition approach is that it enables the training of a simple queueing building block while still supporting the analysis of complex queueing networks. However, applying this method to networks with feedback loops presents a significant challenge, as the building blocks are no longer independent. This issue merits further investigation, since decomposing such networks is inherently difficult, and analyzing them in their entirety is often computationally intensive and inefficient.

\subsection{Limitations}\label{sec:limitations}

Our approach faces two primary limitations. The first is the significant computational effort required for data generation, as it depends on simulation. Although this is feasible for single-station systems, scaling to non-feed-forward networks may render it impractical.

The second limitation is related to extrapolation—making inferences for instances outside the training domain often leads to increased error. For example, applying our approach to a system with 15 servers reduces accuracy when the model was trained on systems with up to 10 servers. While the training domain can be extended to address this, doing so requires frequent retraining, which can be cumbersome.

\section{Conclusion}\label{sec:conclusions}

In this paper, we propose a machine learning-based approach for analyzing multi-server queueing systems. We consider two models: the $GI/GI/c$ and the $GI/GI_i/2$, and aim to predict the steady-state distribution of the number of customers in the system. For the $GI/GI/c$ model, our method achieves state-of-the-art performance when compared to existing results in the literature. Moreover, to the best of our knowledge, our approach is the first to provide an approximation method for the $GI/GI_i/2$ model. Our method demonstrates high accuracy and efficiency across both systems, with errors typically below 3\%, and the ability to evaluate thousands of instances in parallel within a fraction of a second.

Our approximations are based on the first four moments of the inter-arrival and service time distributions. Our analysis indicates that the fifth moment and beyond do not increase accuracy under our settings. Further investigation is required to better understand the moments' effect on the queue dynamics.

Based on our results, further work is needed to extend the ML approach to more complex queueing systems with multi-server queues. This includes systems with blocking, reneging, transient behavior, and non-renewal arrivals—a vital aspect for modeling queueing networks, where upstream station output processes, serving as inputs to downstream stations, are typically non-renewal.







\begin{thebibliography}{30}
\ifx \bisbn   \undefined \def \bisbn  #1{ISBN #1}\fi
\ifx \binits  \undefined \def \binits#1{#1}\fi
\ifx \bauthor  \undefined \def \bauthor#1{#1}\fi
\ifx \batitle  \undefined \def \batitle#1{#1}\fi
\ifx \bjtitle  \undefined \def \bjtitle#1{#1}\fi
\ifx \bvolume  \undefined \def \bvolume#1{\textbf{#1}}\fi
\ifx \byear  \undefined \def \byear#1{#1}\fi
\ifx \bissue  \undefined \def \bissue#1{#1}\fi
\ifx \bfpage  \undefined \def \bfpage#1{#1}\fi
\ifx \blpage  \undefined \def \blpage #1{#1}\fi
\ifx \burl  \undefined \def \burl#1{\textsf{#1}}\fi
\ifx \doiurl  \undefined \def \doiurl#1{\url{https://doi.org/#1}}\fi
\ifx \betal  \undefined \def \betal{\textit{et al.}}\fi
\ifx \binstitute  \undefined \def \binstitute#1{#1}\fi
\ifx \binstitutionaled  \undefined \def \binstitutionaled#1{#1}\fi
\ifx \bctitle  \undefined \def \bctitle#1{#1}\fi
\ifx \beditor  \undefined \def \beditor#1{#1}\fi
\ifx \bpublisher  \undefined \def \bpublisher#1{#1}\fi
\ifx \bbtitle  \undefined \def \bbtitle#1{#1}\fi
\ifx \bedition  \undefined \def \bedition#1{#1}\fi
\ifx \bseriesno  \undefined \def \bseriesno#1{#1}\fi
\ifx \blocation  \undefined \def \blocation#1{#1}\fi
\ifx \bsertitle  \undefined \def \bsertitle#1{#1}\fi
\ifx \bsnm \undefined \def \bsnm#1{#1}\fi
\ifx \bsuffix \undefined \def \bsuffix#1{#1}\fi
\ifx \bparticle \undefined \def \bparticle#1{#1}\fi
\ifx \barticle \undefined \def \barticle#1{#1}\fi
\bibcommenthead
\ifx \bconfdate \undefined \def \bconfdate #1{#1}\fi
\ifx \botherref \undefined \def \botherref #1{#1}\fi
\ifx \url \undefined \def \url#1{\textsf{#1}}\fi
\ifx \bchapter \undefined \def \bchapter#1{#1}\fi
\ifx \bbook \undefined \def \bbook#1{#1}\fi
\ifx \bcomment \undefined \def \bcomment#1{#1}\fi
\ifx \oauthor \undefined \def \oauthor#1{#1}\fi
\ifx \citeauthoryear \undefined \def \citeauthoryear#1{#1}\fi
\ifx \endbibitem  \undefined \def \endbibitem {}\fi
\ifx \bconflocation  \undefined \def \bconflocation#1{#1}\fi
\ifx \arxivurl  \undefined \def \arxivurl#1{\textsf{#1}}\fi
\csname PreBibitemsHook\endcsname

\bibitem[\protect\citeauthoryear{Seelen}{1986}]{SEELEN1986118}
\begin{barticle}
\bauthor{\bsnm{Seelen}, \binits{L.P.}}:
\batitle{An algorithm for ph/ph/c queues}.
\bjtitle{European Journal of Operational Research}
\bvolume{23}(\bissue{1}),
\bfpage{118}--\blpage{127}
(\byear{1986})
\doiurl{10.1016/0377-2217(86)90222-5}
\end{barticle}
\endbibitem

\bibitem[\protect\citeauthoryear{Li and Stanford}{2016}]{LI2016866}
\begin{barticle}
\bauthor{\bsnm{Li}, \binits{N.}},
\bauthor{\bsnm{Stanford}, \binits{D.A.}}:
\batitle{Multi-server accumulating priority queues with heterogeneous servers}.
\bjtitle{European Journal of Operational Research}
\bvolume{252}(\bissue{3}),
\bfpage{866}--\blpage{878}
(\byear{2016})
\doiurl{10.1016/j.ejor.2016.02.010}
\end{barticle}
\endbibitem

\bibitem[\protect\citeauthoryear{Calvo and Arteaga}{2024}]{app14010424}
\begin{botherref}
\oauthor{\bsnm{Calvo}, \binits{R.}},
\oauthor{\bsnm{Arteaga}, \binits{A.}}:
Production systems with parallel heterogeneous servers of limited capacity: Accurate modeling and performance analysis.
Applied Sciences
\textbf{14}(1)
(2024)
\doiurl{10.3390/app14010424}
\end{botherref}
\endbibitem

\bibitem[\protect\citeauthoryear{WHITT}{1993}]{https://doi.org/10.1111/j.1937-5956.1993.tb00094.x}
\begin{barticle}
\bauthor{\bsnm{WHITT}, \binits{W.}}:
\batitle{Approximations for the gi/g/m queue}.
\bjtitle{Production and Operations Management}
\bvolume{2}(\bissue{2}),
\bfpage{114}--\blpage{161}
(\byear{1993})
\doiurl{10.1111/j.1937-5956.1993.tb00094.x}
{\href{https://arxiv.org/abs/https://onlinelibrary.wiley.com/doi/pdf/10.1111/j.1937-5956.1993.tb00094.x}{{https://onlinelibrary.wiley.com/doi/pdf/10.1111/j.1937-5956.1993.tb00094.x}}}
\end{barticle}
\endbibitem

\bibitem[\protect\citeauthoryear{Chaves and Gosavi}{2022}]{Chaves2022}
\begin{barticle}
\bauthor{\bsnm{Chaves}, \binits{C.}},
\bauthor{\bsnm{Gosavi}, \binits{A.}}:
\batitle{On general multi-server queues with non-poisson arrivals and medium traffic: a new approximation and a covid-19 ventilator case study}.
\bjtitle{Operational Research}
\bvolume{22}(\bissue{5}),
\bfpage{5205}--\blpage{5229}
(\byear{2022})
\doiurl{10.1007/s12351-022-00712-2}
\end{barticle}
\endbibitem

\bibitem[\protect\citeauthoryear{Shore}{1988}]{doi:10.1057/jors.1988.45}
\begin{barticle}
\bauthor{\bsnm{Shore}, \binits{H.}}:
\batitle{Simple approximations for the gi/g/c queue—i: The steady-state probabilities}.
\bjtitle{Journal of the Operational Research Society}
\bvolume{39}(\bissue{3}),
\bfpage{279}--\blpage{284}
(\byear{1988})
\doiurl{10.1057/jors.1988.45}
{\href{https://arxiv.org/abs/https://doi.org/10.1057/jors.1988.45}{{https://doi.org/10.1057/jors.1988.45}}}
\end{barticle}
\endbibitem

\bibitem[\protect\citeauthoryear{Gupta et~al.}{2010}]{Gupta2010}
\begin{barticle}
\bauthor{\bsnm{Gupta}, \binits{V.}},
\bauthor{\bsnm{Harchol-Balter}, \binits{M.}},
\bauthor{\bsnm{Dai}, \binits{J.G.}},
\bauthor{\bsnm{Zwart}, \binits{B.}}:
\batitle{On the inapproximability of m/g/k: why two moments of job size distribution are not enough}.
\bjtitle{Queueing Systems}
\bvolume{64}(\bissue{1}),
\bfpage{5}--\blpage{48}
(\byear{2010})
\doiurl{10.1007/s11134-009-9133-x}
\end{barticle}
\endbibitem

\bibitem[\protect\citeauthoryear{Sinreich and Marmor}{2005}]{doi:10.1080/07408170590899625}
\begin{barticle}
\bauthor{\bsnm{Sinreich}, \binits{D.}},
\bauthor{\bsnm{Marmor}, \binits{Y.}}:
\batitle{Emergency department operations: The basis for developing a simulation tool}.
\bjtitle{IIE Transactions}
\bvolume{37}(\bissue{3}),
\bfpage{233}--\blpage{245}
(\byear{2005})
\doiurl{10.1080/07408170590899625}
{\href{https://arxiv.org/abs/https://doi.org/10.1080/07408170590899625}{{https://doi.org/10.1080/07408170590899625}}}
\end{barticle}
\endbibitem

\bibitem[\protect\citeauthoryear{Zeltyn et~al.}{2011}]{10.1145/2000494.2000497}
\begin{botherref}
\oauthor{\bsnm{Zeltyn}, \binits{S.}},
\oauthor{\bsnm{Marmor}, \binits{Y.N.}},
\oauthor{\bsnm{Mandelbaum}, \binits{A.}},
\oauthor{\bsnm{Carmeli}, \binits{B.}},
\oauthor{\bsnm{Greenshpan}, \binits{O.}},
\oauthor{\bsnm{Mesika}, \binits{Y.}},
\oauthor{\bsnm{Wasserkrug}, \binits{S.}},
\oauthor{\bsnm{Vortman}, \binits{P.}},
\oauthor{\bsnm{Shtub}, \binits{A.}},
\oauthor{\bsnm{Lauterman}, \binits{T.}},
\oauthor{\bsnm{Schwartz}, \binits{D.}},
\oauthor{\bsnm{Moskovitch}, \binits{K.}},
\oauthor{\bsnm{Tzafrir}, \binits{S.}},
\oauthor{\bsnm{Basis}, \binits{F.}}:
Simulation-based models of emergency departments: Operational, tactical, and strategic staffing.
ACM Trans. Model. Comput. Simul.
\textbf{21}(4)
(2011)
\doiurl{10.1145/2000494.2000497}
\end{botherref}
\endbibitem

\bibitem[\protect\citeauthoryear{Dieker et~al.}{2017}]{doi:10.1287/opre.2016.1554}
\begin{barticle}
\bauthor{\bsnm{Dieker}, \binits{A.B.}},
\bauthor{\bsnm{Ghosh}, \binits{S.}},
\bauthor{\bsnm{Squillante}, \binits{M.S.}}:
\batitle{Optimal resource capacity management for stochastic networks}.
\bjtitle{Operations Research}
\bvolume{65}(\bissue{1}),
\bfpage{221}--\blpage{241}
(\byear{2017})
\doiurl{10.1287/opre.2016.1554}
{\href{https://arxiv.org/abs/https://doi.org/10.1287/opre.2016.1554}{{https://doi.org/10.1287/opre.2016.1554}}}
\end{barticle}
\endbibitem

\bibitem[\protect\citeauthoryear{Baron et~al.}{2022}]{10015451}
\begin{bchapter}
\bauthor{\bsnm{Baron}, \binits{O.}},
\bauthor{\bsnm{Krass}, \binits{D.}},
\bauthor{\bsnm{Sherzer}, \binits{E.}},
\bauthor{\bsnm{Senderovich}, \binits{A.}}:
\bctitle{Can machines solve general queueing problems?}
In: \bbtitle{2022 Winter Simulation Conference (WSC)},
pp. \bfpage{2830}--\blpage{2841}
(\byear{2022}).
\doiurl{10.1109/WSC57314.2022.10015451}
\end{bchapter}
\endbibitem

\bibitem[\protect\citeauthoryear{Baron et~al.}{2024}]{sherzer23}
\begin{barticle}
\bauthor{\bsnm{Baron}, \binits{O.}},
\bauthor{\bsnm{Krass}, \binits{D.}},
\bauthor{\bsnm{Senderovich}, \binits{A.}},
\bauthor{\bsnm{Sherzer}, \binits{E.}}:
\batitle{Supervised ml for solving the gi/gi/1 queue}.
\bjtitle{INFORMS Journal on Computing}
\bvolume{36}(\bissue{3}),
\bfpage{766}--\blpage{786}
(\byear{2024})
\doiurl{10.1287/ijoc.2022.0263}
\end{barticle}
\endbibitem

\bibitem[\protect\citeauthoryear{Sherzer}{2024}]{sherzer2024computingsteadystateprobabilitiestandem}
\begin{botherref}
\oauthor{\bsnm{Sherzer}, \binits{E.}}:
Computing the steady-state probabilities of a tandem queueing system, a Machine Learning approach
(2024).
\url{https://arxiv.org/abs/2411.07599}
\end{botherref}
\endbibitem

\bibitem[\protect\citeauthoryear{Nii et~al.}{2020}]{Nii20}
\begin{bchapter}
\bauthor{\bsnm{Nii}, \binits{S.}},
\bauthor{\bsnm{Okudal}, \binits{T.}},
\bauthor{\bsnm{Wakita}, \binits{T.}}:
\bctitle{A performance evaluation of queueing systems by machine learning}.
In: \bbtitle{2020 IEEE International Conference on Consumer Electronics - Taiwan (ICCE-Taiwan), Taoyuan, Taiwan},
pp. \bfpage{1}--\blpage{2}
(\byear{2020}).
\doiurl{10.1109/ICCE-Taiwan49838.2020.9258268}
\end{bchapter}
\endbibitem

\bibitem[\protect\citeauthoryear{Efrosinin et~al.}{2025}]{math13050776}
\begin{botherref}
\oauthor{\bsnm{Efrosinin}, \binits{D.}},
\oauthor{\bsnm{Vishnevsky}, \binits{V.}},
\oauthor{\bsnm{Stepanova}, \binits{N.}},
\oauthor{\bsnm{Sztrik}, \binits{J.}}:
Use cases of machine learning in queueing theory based on a gi/g/k system.
Mathematics
\textbf{13}(5)
(2025)
\doiurl{10.3390/math13050776}
\end{botherref}
\endbibitem

\bibitem[\protect\citeauthoryear{Sherzer et~al.}{2024}]{SHERZER2024}
\begin{barticle}
\bauthor{\bsnm{Sherzer}, \binits{E.}},
\bauthor{\bsnm{Baron}, \binits{O.}},
\bauthor{\bsnm{Krass}, \binits{D.}},
\bauthor{\bsnm{Resheff}, \binits{Y.}}:
\batitle{Approximating g(t)/gi/1 queues with deep learning}.
\bjtitle{European Journal of Operational Research}
(\byear{2024})
\doiurl{10.1016/j.ejor.2024.12.030}
\end{barticle}
\endbibitem

\bibitem[\protect\citeauthoryear{Tan and Khayyati}{2022}]{doi:10.1080/00207543.2021.1887536}
\begin{barticle}
\bauthor{\bsnm{Tan}, \binits{B.}},
\bauthor{\bsnm{Khayyati}, \binits{S.}}:
\batitle{Supervised learning-based approximation method for single-server open queueing networks with correlated interarrival and service times}.
\bjtitle{International Journal of Production Research}
\bvolume{60}(\bissue{22}),
\bfpage{6822}--\blpage{6847}
(\byear{2022})
\doiurl{10.1080/00207543.2021.1887536}
{\href{https://arxiv.org/abs/https://doi.org/10.1080/00207543.2021.1887536}{{https://doi.org/10.1080/00207543.2021.1887536}}}
\end{barticle}
\endbibitem

\bibitem[\protect\citeauthoryear{Asmussen}{2003}]{Asmussen2003}
\begin{bbook}
\bauthor{\bsnm{Asmussen}, \binits{S.}}:
\bbtitle{Applied Probability and Queues},
pp. \bfpage{3}--\blpage{38}.
\bpublisher{Springer},
\blocation{New York, NY}
(\byear{2003}).
\doiurl{10.1007/0-387-21525-5_1} .
\burl{https://doi.org/10.1007/0-387-21525-5_1}
\end{bbook}
\endbibitem

\bibitem[\protect\citeauthoryear{Marchal}{1985}]{Marchal1985}
\begin{barticle}
\bauthor{\bsnm{Marchal}, \binits{W.G.}}:
\batitle{Numerical performance of approximate queuing formulae with application to flexible manufacturing systems}.
\bjtitle{Annals of Operations Research}
\bvolume{3}(\bissue{3}),
\bfpage{141}--\blpage{152}
(\byear{1985})
\doiurl{10.1007/BF02024743}
\end{barticle}
\endbibitem

\bibitem[\protect\citeauthoryear{Kraemer~W}{1976}]{KLB76}
\begin{barticle}
\bauthor{\bsnm{Kraemer~W}, \binits{L.-B.M.}}:
\batitle{Approximate formulae for the delay in the queueing system $gi/g/1$ .}
\bjtitle{In: Proceedings of the 8th international telegraphic congress}
\bvolume{2}(\bissue{3}),
\bfpage{231}--\blpage{235}
(\byear{1976})
\end{barticle}
\endbibitem

\bibitem[\protect\citeauthoryear{Marchal}{1976}]{doi:10.1080/05695557608975111}
\begin{barticle}
\bauthor{\bsnm{Marchal}, \binits{W.G.}}:
\batitle{An approximate formula for waiting time in single server queues}.
\bjtitle{A I I E Transactions}
\bvolume{8}(\bissue{4}),
\bfpage{473}--\blpage{474}
(\byear{1976})
\doiurl{10.1080/05695557608975111}
{\href{https://arxiv.org/abs/https://doi.org/10.1080/05695557608975111}{{https://doi.org/10.1080/05695557608975111}}}
\end{barticle}
\endbibitem

\bibitem[\protect\citeauthoryear{Bertsimas}{1990}]{doi:10.1287/opre.38.1.139}
\begin{barticle}
\bauthor{\bsnm{Bertsimas}, \binits{D.}}:
\batitle{An analytic approach to a general class of g/g/s queueing systems}.
\bjtitle{Operations Research}
\bvolume{38}(\bissue{1}),
\bfpage{139}--\blpage{155}
(\byear{1990})
\doiurl{10.1287/opre.38.1.139}
{\href{https://arxiv.org/abs/https://doi.org/10.1287/opre.38.1.139}{{https://doi.org/10.1287/opre.38.1.139}}}
\end{barticle}
\endbibitem

\bibitem[\protect\citeauthoryear{Kimura}{1995}]{KIMURA1995157}
\begin{barticle}
\bauthor{\bsnm{Kimura}, \binits{T.}}:
\batitle{Approximations for the delay probability in the m/g/s queue}.
\bjtitle{Mathematical and Computer Modelling}
\bvolume{22}(\bissue{10}),
\bfpage{157}--\blpage{165}
(\byear{1995})
\doiurl{10.1016/0895-7177(95)00192-5}
\end{barticle}
\endbibitem

\bibitem[\protect\citeauthoryear{Chao and Zhao}{1998}]{CHAO1998392}
\begin{barticle}
\bauthor{\bsnm{Chao}, \binits{X.}},
\bauthor{\bsnm{Zhao}, \binits{Y.Q.}}:
\batitle{Analysis of multi-server queues with station and server vacations}.
\bjtitle{European Journal of Operational Research}
\bvolume{110}(\bissue{2}),
\bfpage{392}--\blpage{406}
(\byear{1998})
\doiurl{10.1016/S0377-2217(97)00253-1} .
\bcomment{EURO Best Applied Paper Competition}
\end{barticle}
\endbibitem

\bibitem[\protect\citeauthoryear{Boxma et~al.}{2002}]{Boxma2002}
\begin{barticle}
\bauthor{\bsnm{Boxma}, \binits{O.J.}},
\bauthor{\bsnm{Deng}, \binits{Q.}},
\bauthor{\bsnm{Zwart}, \binits{A.P.}}:
\batitle{Waiting-time asymptotics for the m/g/2 queue with heterogeneous servers}.
\bjtitle{Queueing Systems}
\bvolume{40}(\bissue{1}),
\bfpage{5}--\blpage{31}
(\byear{2002})
\doiurl{10.1023/A:1017913826973}
\end{barticle}
\endbibitem

\bibitem[\protect\citeauthoryear{Whitt and You}{2022}]{https://doi.org/10.1002/nav.22010}
\begin{barticle}
\bauthor{\bsnm{Whitt}, \binits{W.}},
\bauthor{\bsnm{You}, \binits{W.}}:
\batitle{A robust queueing network analyzer based on indices of dispersion}.
\bjtitle{Naval Research Logistics (NRL)}
\bvolume{69}(\bissue{1}),
\bfpage{36}--\blpage{56}
(\byear{2022})
\doiurl{10.1002/nav.22010}
{\href{https://arxiv.org/abs/https://onlinelibrary.wiley.com/doi/pdf/10.1002/nav.22010}{{https://onlinelibrary.wiley.com/doi/pdf/10.1002/nav.22010}}}
\end{barticle}
\endbibitem

\bibitem[\protect\citeauthoryear{Whitt and You}{2019}]{You19}
\begin{barticle}
\bauthor{\bsnm{Whitt}, \binits{W.}},
\bauthor{\bsnm{You}, \binits{W.}}:
\batitle{A robust queueing network analyzer based on indices of dispersion}.
\bjtitle{Doctor of Philosophy Dissertation}
(\byear{2019})
\doiurl{10.1002/nav.22010}
{\href{https://arxiv.org/abs/https://onlinelibrary.wiley.com/doi/pdf/10.1002/nav.22010}{{https://onlinelibrary.wiley.com/doi/pdf/10.1002/nav.22010}}}
\end{barticle}
\endbibitem

\bibitem[\protect\citeauthoryear{Bandi et~al.}{2015}]{doi:10.1287/opre.2015.1367}
\begin{barticle}
\bauthor{\bsnm{Bandi}, \binits{C.}},
\bauthor{\bsnm{Bertsimas}, \binits{D.}},
\bauthor{\bsnm{Youssef}, \binits{N.}}:
\batitle{Robust queueing theory}.
\bjtitle{Operations Research}
\bvolume{63}(\bissue{3}),
\bfpage{676}--\blpage{700}
(\byear{2015})
\doiurl{10.1287/opre.2015.1367}
{\href{https://arxiv.org/abs/https://doi.org/10.1287/opre.2015.1367}{{https://doi.org/10.1287/opre.2015.1367}}}
\end{barticle}
\endbibitem

\bibitem[\protect\citeauthoryear{Chung et~al.}{2015}]{NIPS2015_b618c321}
\begin{bchapter}
\bauthor{\bsnm{Chung}, \binits{J.}},
\bauthor{\bsnm{Kastner}, \binits{K.}},
\bauthor{\bsnm{Dinh}, \binits{L.}},
\bauthor{\bsnm{Goel}, \binits{K.}},
\bauthor{\bsnm{Courville}, \binits{A.C.}},
\bauthor{\bsnm{Bengio}, \binits{Y.}}:
\bctitle{A recurrent latent variable model for sequential data}.
In: \beditor{\bsnm{Cortes}, \binits{C.}},
\beditor{\bsnm{Lawrence}, \binits{N.}},
\beditor{\bsnm{Lee}, \binits{D.}},
\beditor{\bsnm{Sugiyama}, \binits{M.}},
\beditor{\bsnm{Garnett}, \binits{R.}} (eds.)
\bbtitle{Advances in Neural Information Processing Systems},
vol. \bseriesno{28}.
\bpublisher{Curran Associates, Inc.}, \blocation{???}
(\byear{2015})
\end{bchapter}
\endbibitem

\bibitem[\protect\citeauthoryear{Sherzer et~al.}{2024}]{sherzer2024approximatinggtgi1queuesdeep}
\begin{botherref}
\oauthor{\bsnm{Sherzer}, \binits{E.}},
\oauthor{\bsnm{Baron}, \binits{O.}},
\oauthor{\bsnm{Krass}, \binits{D.}},
\oauthor{\bsnm{Resheff}, \binits{Y.}}:
Approximating G(t)/GI/1 queues with deep learning
(2024).
\url{https://arxiv.org/abs/2407.08765}
\end{botherref}
\endbibitem

\end{thebibliography}

\begin{appendices}

\section{SCV and $\rho$ of Test set (i)}\label{append:scv_rho_testset1}

Here, the distribution of the inter-arrival and service time SCV values along the $\rho$ values under the $GI/GI/1$ system is presented.  

\begin{figure}
    \centering
    \begin{minipage}{0.45\textwidth}
        \centering
        \includegraphics[width=\textwidth]{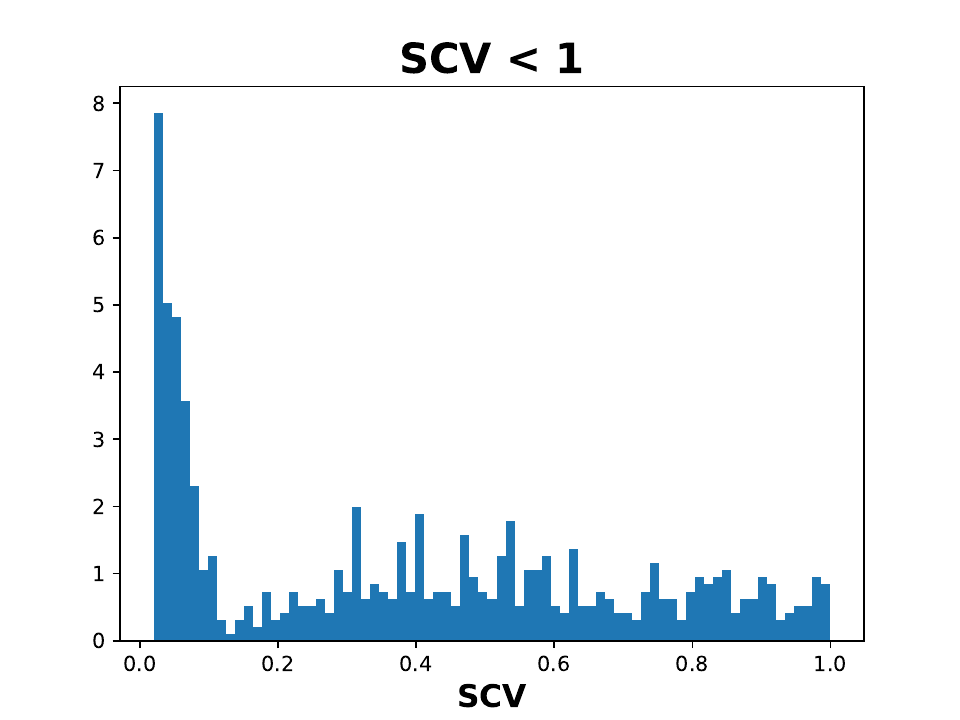}
        \caption{Histogram of low SCV values under Test set (i)}
        \label{fig:low_scv}
    \end{minipage}
    \hfill
    \begin{minipage}{0.45\textwidth}
        \centering
        \includegraphics[width=\textwidth]{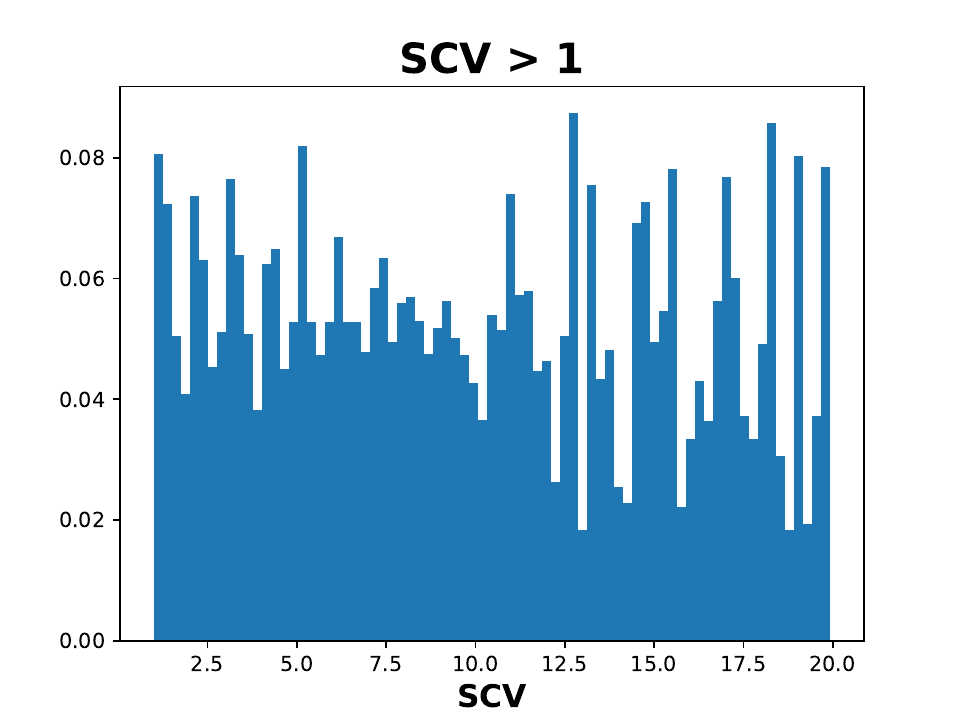}
        \caption{Histogram of high SCV values under Test set (i)}
        \label{fig:high_scv}
    \end{minipage}
\end{figure}

\begin{figure}
\centering
\includegraphics[scale=0.4]{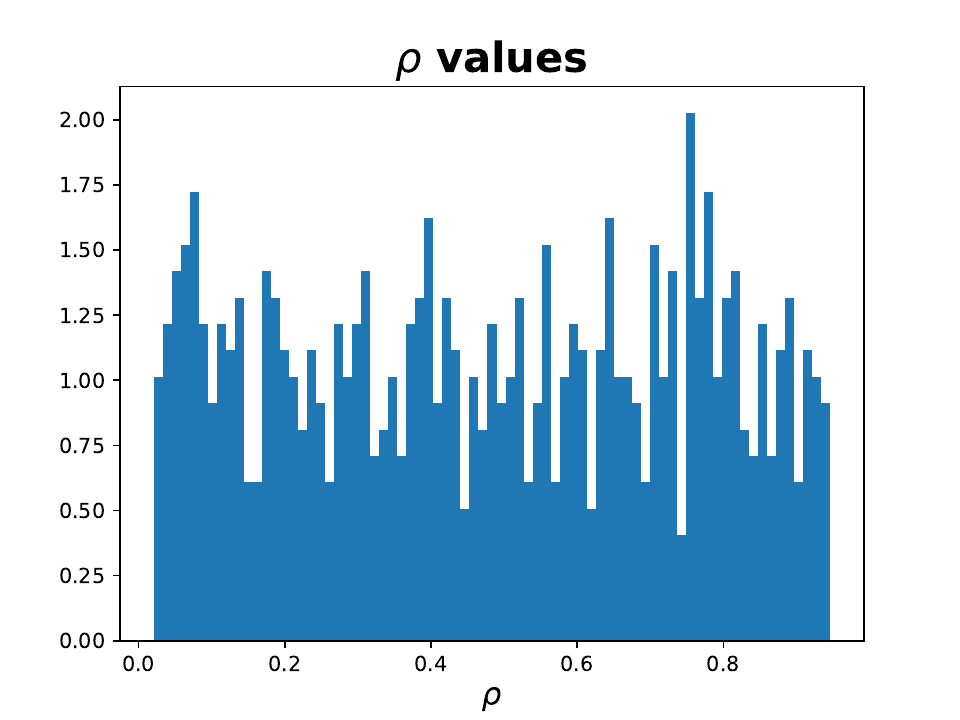}
\caption{Histrogram of $\rho$ under Test set (i) }
\label{fig:rhos}
\end{figure}




\end{appendices}



\end{document}